\documentclass[iop,apj,numberedappendix,appendixfloats,twocolappendix]{emulateapj}

\shorttitle{\sc The Vulture Survey I}
\shortauthors{\sc Mathes et~al.}

\usepackage{natbib}
\usepackage{iondefs}
\usepackage{apjfonts}
\usepackage{mathtools}
\usepackage{amsmath}
\newcommand{\angstrom}{\textup{\AA}}

\begin{document}

\title{The Vulture Survey I: Analyzing the Evolution of ${\MgII}$ Absorbers}

\author{
Nigel L. Mathes\altaffilmark{1},
Christopher W. Churchill\altaffilmark{1},
and
Michael T. Murphy\altaffilmark{2}
}

\altaffiltext{1}{New Mexico State University, Las Cruces, NM 88003, United States}
\altaffiltext{2}{Centre for Astrophysics and Supercomputing, Swinburne University of Technology, Victoria 3122, Australia}

\begin{abstract}
We present detailed measurements of the redshift path density, equivalent width distribution, column density distribution, and redshift evolution of ${\MgII}$ absorbers as measured in archival spectra from the UVES spectrograph at the Very Large Telescope (VLT/UVES) and the HIRES spectrograph at the Keck Telescope (Keck/HIRES) to equivalent width detection limits below $0.01$~{\AA}. This survey examines 432 VLT/UVES spectra from the UVES SQUAD collaboration and 170 Keck/HIRES spectra from the KODIAQ group, allowing for detections of intervening ${\MgII}$ absorbers spanning redshifts $0.1 < z < 2.6$. We employ an accurate, automated approach to line detection which consistently detects redshifted absorption lines. We measure the equivalent widths, apparent optical depth column densities, and velocity widths for each absorbing system. Using our complete sample of all detectable ${\MgII}$ absorbers, we can accurately determine the redshift path density of absorbers across cosmic time. We measure evolution in the comoving ${\MgII}$ line density, $dN\,/dX$, finding more high equivalent width absorbers at $z = 2$ than at present. We also measure evolution in the equivalent width distribution, parameterized by a Schechter function fit, finding a shallower weak-end slope ($\alpha$) for absorbers at redshifts between $1.53 < z < 2.64$, with $\alpha = -0.81 \pm 0.12$, compared to absorbers between $0.14 < z < 0.78$, where $\alpha = -1.09 \pm 0.09$. Finally, we calculate the cosmic mass fraction of ${\MgII}$ using the column density distribution, finding that $\Omega_{\hbox{\scriptsize {\MgII}}}$ increases from $\Omega_{\hbox{\scriptsize {\MgII}}} = (0.9 \pm 0.2) \times 10^{-8}$ at $<z>\,= 0.49$ to $\Omega_{\hbox{\scriptsize {\MgII}}} = (1.4 \pm 0.2) \times 10^{-8}$ at $<z>\,= 2.1$. We find that weak ${\MgII}$ absorbers, those with equivalent widths less than $0.3$~{\AA}, are physically distinct and evolve separately from very strong ${\MgII}$ absorbers, which have equivalent widths greater than $1.0$~{\AA}. We compare our observed evolutionary trends in the distributions of ${\MgII}$ absorbers to previously studied cosmic trends in metallicty, the ionizing ultraviolet background, and star formation in order to conclude that galaxies eject more metal enriched gas into their halos around $z = 2$ than at lower redshifts in the form of high equivalent width ${\MgII}$-absorbing outflows. Over time from $z = 2$ to the present, these feedback processes decline and the evolving conditions in the circumgalactic medium give rise to a population of low equivalent width, passive ${\MgII}$ absorbers.

\end{abstract}


%============== INTRODUCTION =============================

\section{Introduction}
\label{sec:intro}

One of the most important questions in modern studies of galactic evolution asks, how do baryons cycle into and out of galaxies, and how does this cycle determine the growth and evolution of galaxies themselves? More specifically, how does the process of gas accretion, star formation, and subsequent supernovae-driven feedback shape both the galaxies themselves and their circumgalactic medium (CGM)? By using spectroscopic observations of quasars, we can identify and analyze metal line absorbers in and around the halos of foreground galaxies. Though absorption line studies by themselves cannot directly answer these questions, the statistical results from such studies can provide vital information from which further progress can be made.

One of the most prolific absorption features, the {\MgIIdblt} doublet, traces cool ($T \simeq 10^4~\mathrm{K}$; \cite{Churchill2003}) metal enriched gas in the disks and halos of galaxies. It is one of the best tracers of this gas because it can exist in a wide range of ionizing conditions, ranging in ionization parameter from $-5 < \log U < 1$~\citep{Churchill1999}, it is observable in optical wavelengths for redshifts between $0.1 < z < 2.6$, and it has predictable line characteristics defined by its resonant doublet nature which make it ideal for automated searches.

The origin of ${\MgII}$ absorbing gas is still debated. As summarized in~\cite{Kacprzak2011} and~\cite{Matejek2013}, two separate interpretations exist to explain the origin of strong, high equivalent width ($W_r$) {\MgII} absorbers ($W_r^{\lambda2796} > 0.3$~{\AA}) and weak, low equivalent width ($W_r^{\lambda2796} < 0.3$~{\AA}) ${\MgII}$ absorbers. For the strong, higher equivalent width systems, multiple correlations exist between the rest frame ${\MgII}$ equivalent width around galaxies and the host galaxy's star formation properties. \cite{Zibetti2007}, \cite{Lundgren2009}, \cite{Noterdaeme2010}, \cite{Bordoloi2011}, and \cite{Nestor2011} all found a correlation betwen $W_r^{\lambda2796}$ and blue host galaxy color, showing that galaxies with more active star formation have more metal enriched gas in their halos. \cite{Bordoloi2014} also found that ${\MgII}$ equivalent width increases with increasing star formation rate density. In addition, spectroscopic observations of star forming galaxies have revealed strong ${\MgII}$ absorption blueshifted $300 - 1000$~{\kms} relative to the host galaxy~\citep{Tremonti2007,Weiner2009,Martin2009,Rubin2010}. \cite{Bouche2006} found an anti-correlation between galaxy halo mass, derived from the cross-correlation between ${\MgII}$ absorption systems and luminous red galaxies, and ${\MgII}$ equivalent width, showing that individual clouds of a ${\MgII}$ system are not virialized in the halos of galaxies. They interpreted their results as a strong indication that high equivalent width absorbers with $W_r^{\lambda2796} \gtrsim 2$~{\AA} arise in galactic outflows. Marginal anti-correlations between ${\MgII}$ equivalent width and galaxy halo mass using the same cross-correlation method were also reported by~\cite{Gauthier2009} and~\cite{Lundgren2009}. It is imporant to note, however, that~\cite{Churchill2013letter} and~\cite{MAGIICAT3} find no correlation between $W_r^{\lambda2796}$ and halo mass when halo mass is derived from abundance matching. They instead find that galaxies inhabiting more massive dark matter halos have stronger absorption at a given distance.

For the weak, lower equivalent width systems, it seems none of the above correlations hold. \cite{Chen2010b}, \cite{Kacprzak2011}, and \cite{Lovegrove2011} found little evidence for a correlation between galaxy color and ${\MgII}$ equivalent width when restricting their samples to weak absorbers. \cite{Kacprzak2011} measured the orientation of galaxies relative to ${\MgII}$ detections in the sight lines of background quasars and identified low metallicity, low equivalent width ${\MgII}$ absorbers co-planar with some some galaxy disks, implying structures associated with accreting filaments as opposed to outflows, which are more often observed perpendicular to the galaxy disk~\citep{Bordoloi2011,Kacprzak2012-PA}. Finally, the simulations of~\cite{Stewart2011} and~\cite{Ford2013mass} revealed a reservoir of low-ionization, metal enriched, co-rotating gas around massive galaxies. Together, these studies imply that weak ${\MgII}$ absorption systems may preferentially trace low metallicity infall and co-rotating gas in the circumgalactic medium.

\cite{MAGIICAT1} constructed a sample of ${\MgII}$ absorbers and their associated galaxies and examined both strong and weak ${\MgII}$ absorbers from $0.07 \le z \le 1.1$. In the subsequent analysis of their sample,~\cite{MAGIICAT2} found a more extended ${\MgII}$ absorbing CGM around higher luminosity, bluer, higher redshift galaxies. In addition, in~\cite{MAGIICAT4}, they found that bluer galaxies replenish their ${\MgII}$ absorbing CGM through outflows, whereas red galaxies do not. Finally, in~\cite{MAGIICAT5}, it is made clear that the largest velocity dispersions in ${\MgII}$ absorbing systems are measured around blue, face-on galaxies probed along their minor axis, strongly suggesting that these ${\MgII}$ absorbers originate in bi-conical outflows.

Many surveys have been undertaken to inventory ${\MgII}$ absorbers and examine their evolution. The earliest studies~\citep{Lanzetta1987,Tytler1987,Sargent1988,Steidel1992} found that ${\MgII}$ systems with rest equivalent widths above $0.3$~{\AA} show no evolution in $dN\!/dz$ between redshifts $0.2 < z < 2.15$. These studies also found that the equivalent width distribution function, $f(W_r^{\lambda2796})$, could be fit equally well with either an exponential or a power-law function. It remains uncertain whether the cosmic distribution of ${\MgII}$ in galactic halos exhibits a fractal, self-similar nature, or if $f(W_r^{\lambda2796})$ flattens at equivalent widths below $W_r^{\lambda2796} < 0.3$~{\AA}.

${\MgII}$ absorption surveys have taken one of two different approaches to try to analyze the global distribution of ${\MgII}$ absorbing gas across cosmic time. \cite{Churchill1999} and \cite{Narayanan2007} aimed to determine more precisely how $dN\!/dz$ and $f(W_r^{\lambda2796})$ evolve with redshift by surveying weak ${\MgII}$ absorbers. They found that, for these low equivalent width absorbers, $dN\!/dz$ increases as a function of increasing redshift up until $z = 1.4$. At higher redshifts, $dN\!/dz$ falls to lower values, though the uncertainties are large. In addition, they found the equivalent width distribution function for weak absorbers is best fit by a power-law, strongly disfavoring an exponential fit to the overall distribution.

The most recent studies have employed new multi-object spectrographs such as the Sloan Digital Sky Survey (SDSS) and the FIRE spectrograph on the Magellan Baade Telescope~\citep{Nestor2005,Matejek2012,Chen2016}. \cite{Nestor2005}, who examined over 1300 intervening ${\MgII}$ absorbers in SDSS quasar spectra with $W_r^{\lambda2796} > 0.3$~{\AA}, found that the equivalent width distribution function is well fit by an exponential. They did not find evidence for redshift evolution in systems with $0.4 < W_r^{\lambda2796} < 2$~{\AA}, but observed an enhancement in the number of $W_r^{\lambda2796} > 2$~{\AA} absorbers per comoving redshift path length as a function of increasing redshift from $z \sim 0$ up to $z \sim 2$. \cite{Matejek2012} and~\cite{Chen2016}, analyzing 279 ${\MgII}$ absorbing systems from $2 < z < 7$ in infrared FIRE spectra, also found that the equivalent width distribution function is well fit by an exponential. They also observed that systems with $W_r^{\lambda2796} < 1.0$~{\AA} show no evolution with redshift, but higher equivalent width systems grow in number density from low redshift to $z \sim 3$, after which the number density declines. Collectively, these surveys imply physical changes in the astrophysical processes or in the state of the gas structures in the environments giving rise to ${\MgII}$ absorption as the universe ages.

The properties of a given ${\MgII}$ absorbing cloud are governed by the total amount of gas present, the gas phase metallicity, and the nature of the background radiation incident on the cloud. As shown by~\cite{Quiret2016}, studying a large sample of damped Ly$\alpha$ absorbers (DLAs; neutral hydrogen absorbers with $\log(N(\mathrm{HI})) > 20.3$) and sub-DLAs ($19.0~\mathrm{cm^{-2}} < \log(N(\mathrm{HI})) < 20.3~\mathrm{cm^{-2}}$), the average metallicity of the circumgalactic medium decreases from $z = 0$ to $z = 5$. In addition,~\cite{Menard2009} show that a correlation exists between the neutral hydrogen column density of an absorber and the rest frame ${\MgII}$ equivalent width, with stronger ${\MgII}$ absorbing systems having large ${\HI}$ column densities. As ${\MgII}$ absorbers are often found associated with nearby galaxies, one can expect that the metallicity evolution of DLAs and sub-DLAs should be reflected in the evolution of ${\MgII}$ absorbing systems~\citep{Kulkarni2002,Prochaska2003,Kulkarni2005,Kulkarni2007}.

%Finally, multiple studies have shown that ${\MgII}$ absorption is observed primarily around galaxies within a projected radius of $\sim100~\mathrm{kpc}$.

%using detailed radiative transfer models, synthesized the intensity and shape of the diffuse cosmic UV/X-ray background. They\

The cosmic ionizing background also changes dramatically from $z = 0$ to $z = 2.5$. \cite{Haardt2012} showed that the slope and intensity of the diffuse UV/X-ray ionizing background increase as redshift increases, with a harder, more intense ionizing background present at higher redshift. Specifically, the comoving 1 Rydberg emissivity increases from $\sim2 \times 10^{23}~\mathrm{erg~s^{-1}~Mpc^{-3}~Hz^{-1}}$ at $z = 0$ to $\sim70 \times 10^{23}~\mathrm{erg~s^{-1}~Mpc^{-3}~Hz^{-1}}$ at $z = 2.5$. ${\MgII}$ absorbers are subject primarily to this UV background, with very little contribution from stellar radiation from a nearby galaxy~\citep{Churchill1999,Charlton2000,Rigby2002}.

\cite{Behroozi2013sfr} showed that the cosmic star formation rate peaks around $z \sim 2$. At this point in time, galaxies are on average forming stars at a rate ten times greater than at $z = 0$. In conjunction with the fact that galactic-scale outflows can be driven by star formation~\citep{Zhu2015}, and these outflows can eject ${\MgII}$ absorbing gas to large galactocentric radii~\citep{Sharma2013,Kacprzak2012-PA,Nestor2011}, it follows that more metal enriched gas, traced by ${\MgII}$ absorbers, should be driven out of galaxies at $z \sim 2$. One goal of The Vulture Survey is to find and analyze definitive observational signatures of this energetic epoch.

We now aim to better understand the complex relationship between absorbing gas in the CGM/IGM and the physical processes shaping galaxy formation as the universe ages. For our survey, we will analyze the largest, most comprehensive sample of high resolution, high $S\,/N$ quasar spectra to uniformly observe both strong and weak ${\MgII}$ absorbers. We hope to finally bridge the equivalent width dichotomy in prior ${\MgII}$ absorption line surveys by analyzing large numbers of both strong and weak absorbers. To do so, we will examine quasar spectra observed with either the VLT/UVES~\citep{Dekker2000} or Keck/HIRES~\citep{Vogt1994} spectrographs. We aim to characterize the evolution in the number density of all ${\MgII}$ absorbers from present to beyond the peak of the cosmic star formation rate. We interpret these results in the context of global evolution in metallicity around galaxies, the ionizing background, and cosmic star formation.

We begin by explaining the methods of acquiring and analyzing the quasar spectra in Section~\ref{sec:data}. In Section~\ref{sec:results}, we present the results showing the evolution of the ${\MgII}$ equivalent width distribution, $dN\!/dX$, and the ${\MgII}$ column density distribution across redshift. We also analyze the functional fit to both the equivalent width and column density distributions. In Section~\ref{sec:discussion} we discuss the redshift evolution of all types of ${\MgII}$ absorbers and derive the relative matter density contributed to the universe by {\MgII}, $\Omega_{\hbox{\scriptsize {\MgII}}}$. In Section~\ref{sec:conclusions} we summarize our results and look to future studies using this rich data set, including a companion analysis of intervening {\CIV} absorbers and detailed kinematic analysis of intervening absorbing systems. For all calculations, we adopt the most recently published Planck cosmology, with $H_0 = 67.81~\mathrm{km~s^{-1}~Mpc}$, $\Omega_M = 0.308$, and $\Omega_{\Lambda} = 0.692$~\citep{Planck2016}.

%============== SAMPLE DESCRIPTION, DATA, ANALYSIS =======================

\section{Data and Analysis}
\label{sec:data}

\subsection{Quasar Spectra Sample}

We have assembled a sample of 602 archival quasar spectra observed with the VLT/UVES and Keck/HIRES spectrographs. The data originate from two archival data mining efforts - the UVES SQUAD collaboration (432 spectra) led by Michael Murphy, and the KODIAQ Survey (170 spectra) led by John O'Meara \citep{OMeara2015}. The spectra range in signal-to-noise ratio $(S\,/N)$ from $4$ to $288$ per $1.3 - 2.5~\mathrm{\kms}$ pixel, with the pixel size dependent upon the resolution of the spectrum. The mean $S\,/N$ for the sample is $38$ per pixel. Quasar emission redshifts span $0.014 < z < 5.292$. Wavelength coverage for each spectrum varies based upon the settings used for each spectrograph. VLT/UVES has 3 CCD chips available, offering large wavelength coverage from $\sim 3000$ to $\sim10\,000$~{\AA}. However, wavelength coverage available for each quasar spectum varies based upon the selected cross-disperser settings. The exposures used from Keck/HIRES were taken from 2004 to present, when a 3 chip CCD mosiac was installed, also allowing wavelength coverage from  $\sim 3000$ to $\sim10\,000$~{\AA}. Again, though, individual quasar observations vary in wavelength coverage based upon cross-disperser angle. We detect 1180 ${\MgII}$ absorbing systems from $0.14 < z < 2.64$ to a detection limit of $W_r^{\lambda2796} \simeq 0.01$~{\AA} for regions with $S\,/N > 40$ per pixel.

\begin{figure*}[bth]
\epsscale{1.17}
\plottwo{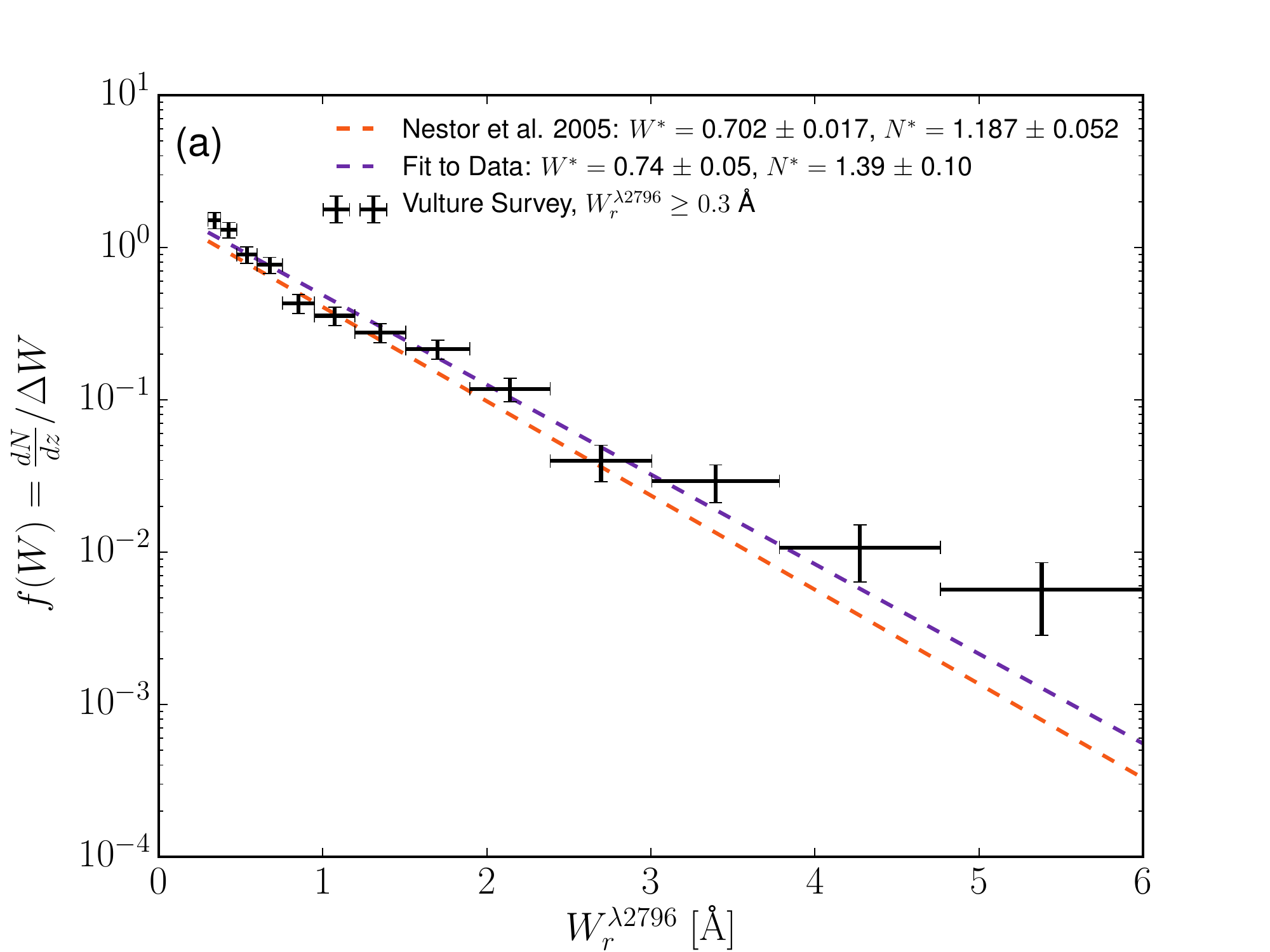}{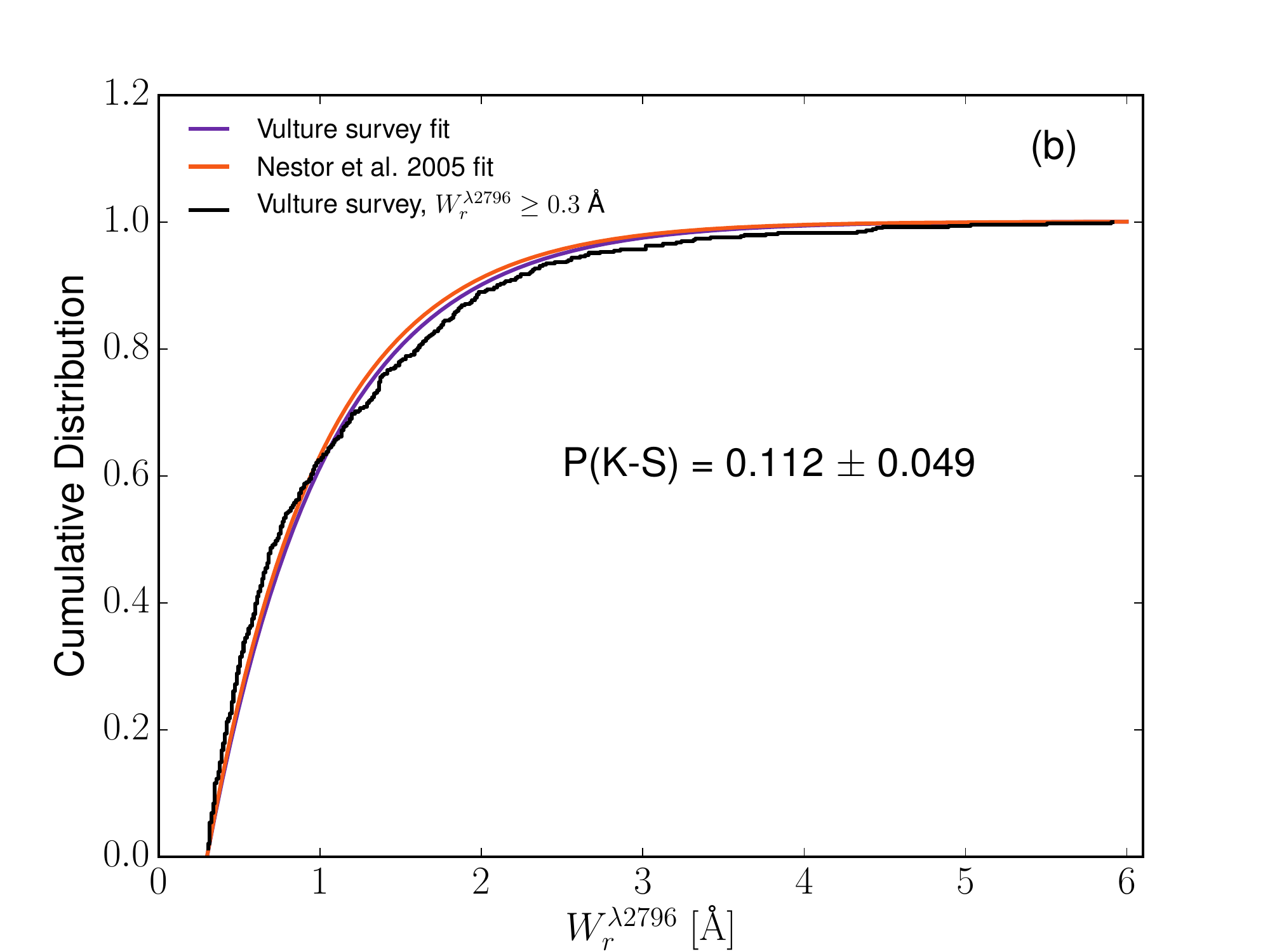}
\caption{(a) The $W_r^{\lambda2796} \ge 0.3$~{\AA} equivalent width frequency distribution for The Vulture Survey in black compared to the exponential fit of~\cite{Nestor2005}, shown as the orange dashed line, and an exponential fit to The Vulture Survey data in purple. The exponential fits are of the form shown in Equation~\ref{eqn:exponential}. (b) The cumulative distribution of The Vulture Survey data and two comparative exponential fits. The P(K-S) value shown compares our survey's data to the fit from~\cite{Nestor2005}. }
\label{fig:bias}
\end{figure*}

The archival quasar spectra used to construct The Vulture Survey were observed for a multitude of reasons. In cases where observations were taken to target a previously known absorption line system, these detections could bias our sample when calculating $dN\!/dz$ and $dN\!/dX$. Because early studies which discovered ${\MgII}$ absorbers did not have the sensitivity to detect weak, low equivalent width systems, quasar spectra were never selected based upon the presence of weak absorption. However, some spectra were selected based on the presence of strong, $W_r^{\lambda2796} > 0.3$~{\AA} systems. In order to properly quantify any bias, we compare our work to the massive, unbiased sample of the Sloan Digital Sky Survey (SDSS). Specifically, we turn to the work of~\cite{Nestor2005} who inventoried strong ${\MgII}$ absorbers in SDSS quasar spectra. In~\cite{Nestor2005}, the authors constructed the equivalent width frequency distribution for absorbers with $W_r^{\lambda2796} > 0.3$~{\AA} and found the best fit to this distribution was an exponential function of the form,

\begin{equation}
f(W_r^{\lambda2796}) = \frac{N^*}{W^*} e^{-\,\left(W_r^{\lambda2796} / W^*\right)}
\label{eqn:exponential}
\end{equation}

\noindent where $N^*$ and $W^*$ are constants. In order to compare to the SDSS data, we limit our sample to absorbers with equivalent widths $W_r^{\lambda2796} > 0.3$~{\AA} and calculate the equivalent width frequency distribution. In Figure~\ref{fig:bias}(a), we show $f(W_r^{\lambda2796})$ for The Vulture Survey and an exponential fit to our data, along with the exponential fit of~\cite{Nestor2005}. \cite{Nestor2005} found the best-fit parameters and corresponding $1\sigma$ uncertainties to be $N^* = 1.187 \pm 0.052$ and $W^* = 0.702 \pm 0.017$. When we perform the same analysis, fitting Equation~\ref{eqn:exponential} to The Vulture Survey data, we derive $N^* = 1.39 \pm 0.10$ and $W^* = 0.74 \pm 0.05$. By eye, there appears a slight excess in the number of ${\MgII}$ absorbers above $W_r^{\lambda2796} > 3$~{\AA}.

In order to statistically determine the bias in our sample, we perform a Kolmogorov--Smirnov (KS) test to quantitatively measure the similarity between our sample of absorbers with $W_r^{\lambda2796} > 0.3$~{\AA} and the SDSS sample of~\cite{Nestor2005}. We first sample a population of 1331 measured absorber equivalent widths, matching the SDSS sample size, from the exponential fit to the equivalent width ditribution of~\cite{Nestor2005}, incorporating the reported $1\sigma$ scatter in the fit parameters. This allows us to directly compare to the sample of measured ${\MgII}$ equivalent widths from The Vulture Survey. In Figure~\ref{fig:bias}(b), we show the cumulative distribution of the strong absorbers in The Vulture Survey in black, along with the exponential fit to our data and the fit of~\cite{Nestor2005}. We then perform a two-sample KS test, calculating the P-value, which is the probability that the two samples are drawn from different underlying distributions, defined as P(K-S). To avoid issues due to the random resampling of the~\cite{Nestor2005} distribution, we repeat this exercise one million times as a Monte-Carlo method, generating an ensemble of P(K-S) values. We take the mean P(K-S) value as the true probability that the samples are inconsistent, and the standard deviation about this mean as the $1\sigma$ uncertainty. Our criterion to assert that our sample is not inconsistent with an unbiased sample requires P(K-S)$\, > 0.0027$, which means that it could not be ruled out at the $3\sigma$ level that the two populations are consistent with one another. With P(K-S)$\, = 0.112 \pm 0.049$, and only one instance out of the million Monte-Carlo runs exhibiting a P(K-S) value below 0.0027, we conclude that our sample is not inconsistent with an unbiased sample, even at the $2\sigma$ level, and that the strong absorbers in The Vulture Survey could very well have originated from the same underlying population as the unbiased SDSS quasar spectra sample.

% ============== Continuum Fitting and Line Detection ======================================

\subsection{Data Reduction and Line Detection}
\label{sec:detection}

The KODIAQ data sample is reduced and fully continuum fit, delivered as normalized spectra according to the prescriptions of \cite{OMeara2015}. To summarize, observing runs are grouped together and uniformly reduced using HIRedux\footnote{http://www.ucolick.org/~xavier/HIRedux/} as part of the XIDL\footnote{http://www.ucolick.org/~xavier/IDL/index.html} suite of astronomical routines in IDL. Continuum fits are applied one order at a time using Legendre polynomials by a single member of the KODIAQ team, John O'Meara, to minimize bias and variation.

The UVES SQUAD sample also comes reduced and continuum fit according to the prescriptions of~\cite{King2012,Bagdonaite2014,Murphy2016,Murphyprep}. Reduction was carried out using the ESO Common Pipeline Language data-reduction software.\footnote{http://www.eso.org/observing/dfo/quality/UVES/pipeline/pipe\_reduc.html} The continuum is fit automatically with a low order polynomial in small sections using UVES\_popler, an ESO/VLT UVES post-pipeline echelle reduction program written by Michael T. Murphy~\citep{MurphyPOPLER}. This fit can incorrectly estimate the continuum around narrow emission regions and broad absorption features. Using UVES\_popler, we add a higher order continuum fit to such regions of each spectrum, always preserving the continuity of the continuum with non-absorbing regions.

The next step involves detecting all ${\MgII}$ absorption features. We first limit the search range to regions of the spectrum redward of the {\Lya} emission, as {\Lya} forest contamination would render automatic detection of weaker metal lines nearly impossible. We also do not search $5000$~{\kms} blueward of the quasar emission redshift in order to avoid absorbers associated with the quasar itself. Finally, we exclude regions of strong telluric absorption bands, specifically from $6277 - 6318$~{\AA}, $6868 - 6932$~{\AA}, $7594 - 7700$~{\AA}, and $9300 - 9630$~{\AA}, because we found that the molecular line separations and ratios can lead to numerous false positives when searching for ${\MgII}$ doublets.

To find all intervening {\MgIIdblt} absorbers, we employ a techinque outlined in \cite{Zhu2013}, in which we perform a matched filter search for absorption candidates detected above a certain S\,/N threshold. The filter is a top hat function centered at the wavelength of the desired redshifted absorption line. Its width is selected to match the resolution of the spectrum, which is a function of the slit width used during the exposure. A large variety of slit-widths were used to achieve different resolutions for varying science drivers, but, characteristically for a 1.0 arcsec slit, $R \sim 40,000$ for VLT/UVES and $R \sim 45,000$ for Keck/HIRES. We convolve the filter with the normalized spectrum to generate a normalized power spectrum in redshift space, with absorption features having positive power.

The error spectrum in both instruments is complex, irregular, and has frequent single-pixel spikes which makes uniform normalization impossible. Therefore, we cannot convolve the filter with the error spectrum to derive normalized noise estimates, as is often done in matched filter analysis. Instead, we examine the noise in the derived power spectrum. To derive the noise, we first sigma-clip chunks of the power spectrum to remove absorption features, leaving only the continuum power spectrum. Next, we calculate the standard deviation of this continuum. Finally, we use the standard deviation as the noise to calculate the $S\,/N$ of the absorption features in the power spectrum as the ratio of the normalized power ($S$) to the normalized noise ($N$). A flagged absorption feature has $S\,/N > 5$. A confirmed doublet detetection for {\MgIIdblt} requires detection of $S\,/N^{\lambda2796} > 5$ and $S\,/N^{\lambda2803} > 3$. In addition, our automated routines remove detections with non-physical doublet ratios in unsaturated regions; specifically, we exclude cases where $W_r^{\lambda2803} > W_r^{\lambda2796}$, or $W_r^{\lambda2803} < \left(0.3 \times W_r^{\lambda2796}\right)$. The latter constraint is conservative for unsaturated ${\MgII}$ absorbers, as $W_r^{\lambda2803}$ is rarely observed less than $0.5 \times W_r^{\lambda2796}$. We relax this constraint in saturated features. This system could potentially exclude detections where either the ${\MgII2796}$ or ${\MgII2803}$ line is blended with another transition but does not saturate; however, confirmation of these cases requires extra verification from separate absorbing features, such as ${\FeII}$, which are weaker and not always covered in the spectrum.

All absorption features are visually verified upon completion of the detection algorithm. Multiple feature detections within $\pm 500$~{\kms} of each other are grouped together to generate absorption systems, designated as a single absorber, to be analyzed. Once absorption systems are identified, we calculate the optical depth-weighted median absorption redshift to define the center of the entire absorption system. The formal derivation of this redshift is described in the appendix of~\cite{Churchill2001}.

We also derive an equivalent width detection limit across the spectrum. To do so, we insert modelled Gaussian absorption features across the spectrum and assume a full-width at half maximum (FWHM) defined by the resolution of the instrument to represent unresolved lines. We then solve for the height of the Gaussian, defined as the value at the curve's peak, required to detect the unresolved line with our matched filtering technique at a $S\,/N = 5$. Finally, we integrate to find the equivalent width, and take that value as the minimum detectable equivalent width at a given wavelength. The detection algorithm is therefore self-monitoring. This full equivalent width detection limit spectrum also allows us to accurately characterize the completeness of our sample, along with the full redshift path length searched.

% ============== Measuring Absorption Properties ======================================

\subsection{Measuring Absorption Properties}
\label{sec:measuring}

For each absorption system, we automatically define the wavelength bounds of an absorbing region by finding where the flux recovers to within $1\sigma$ of the continuum, with $\sigma$ defined by the error spectrum, for three pixels on either side of the absorption trough. Within these regions we calculate rest-frame equivalent widths ($W_r$), velocity widths ($\Delta v$), optical depth-weighted kinematic spreads ($\omega_v$), apparent optical depth (AOD) column densities ($\log(N)$), and absorption asymmetries. The functional forms of these parameters are detailed in the appendix of~\cite{Churchill2001}, equations$~\mathrm{A3 - A7}$.

%============= RESULTS: Basic Absorption Properties =========================

\section{Results}
\label{sec:results}

% ================ Parameter Distributions ================
\subsection{Sample Characterization}
\label{sec:sample}

\begin{figure*}[bth]
\epsscale{1.27}
\plotone{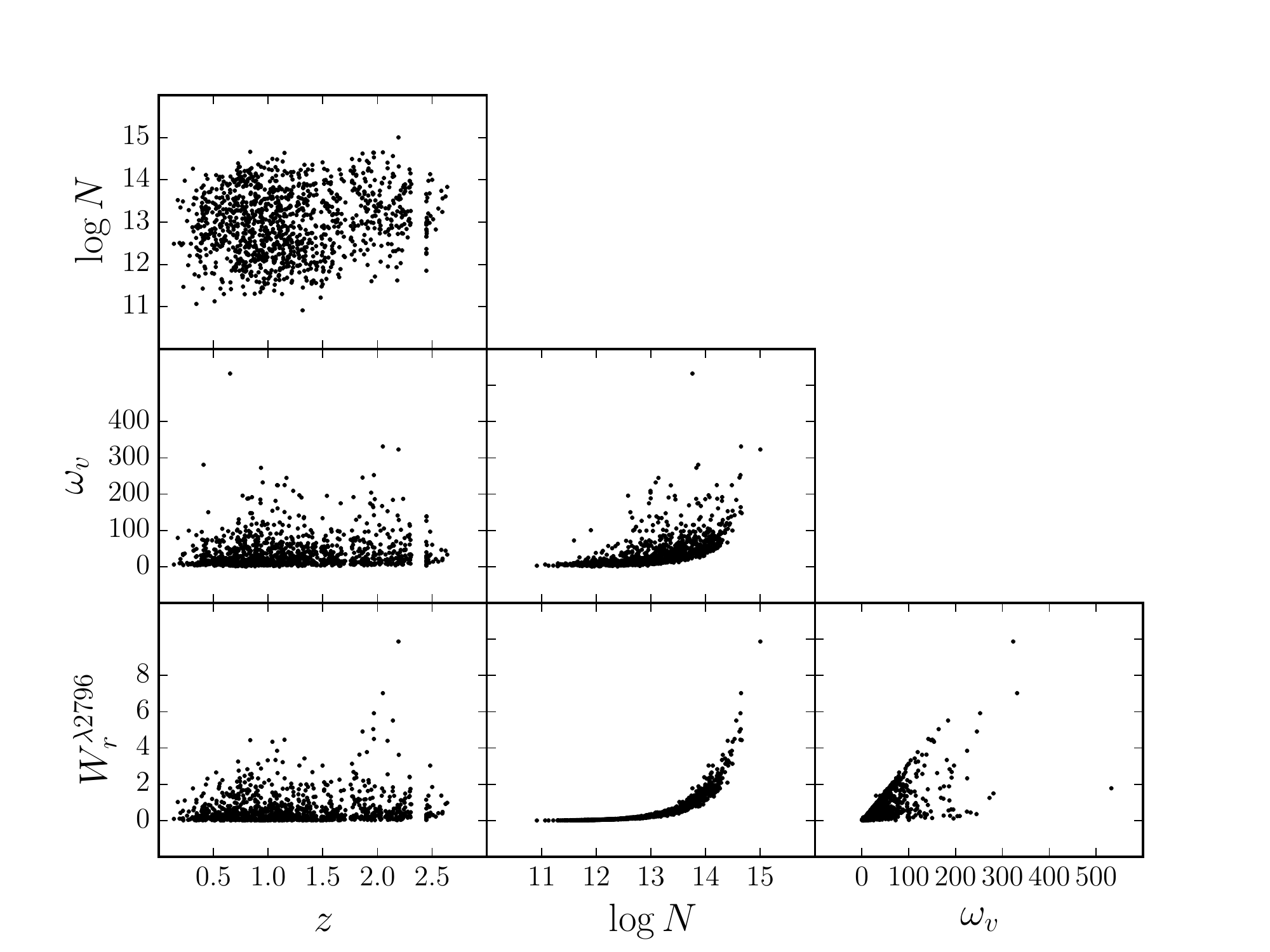}
\caption{Correlations between measured absorption properties for The Vulture Survey. $\log N$ is the ${\MgII}$ AOD column density, $\omega_v$ is the kinematic spread, $W_r^{2796}$ is the rest frame {\MgII2796} equivalent width, and $z$ is the absorption redshift.}
\label{fig:scatterplots}
\end{figure*}

Figure~\ref{fig:scatterplots} shows the relationships between the measured absorption parameters, characterizing the distribution of absorption properties for our survey. With redshift, there are no obvious trends other than the highest equivalent width absorbers, with $W_r^{2796} > 4$~{\AA}, existing mainly at $z > 1.5$. The data gaps at $z = 1.7$ and $z = 2.4$ represent the larger omitted search regions which overlap with the stronger telluric absorption bands. With column density, we see the normal trends of higher column density systems exhibiting higher equivalent widths and velocity spreads, with the distributions asymptoting near $\log N \simeq 15~\mathrm{cm^{-2}}$ due to saturation effects and the nature of measuring column densities with the AOD method. Measured column densitites of saturated lines are lower limits. With respect to kinematic spread, we observe the saturation line in the $\omega_v$ vs. $W_r^{2796}$ relationship, showing the maximum $\omega_v$ for a flat-bottomed absorption profile of a given equivalent width.

\subsection{Sample Completeness and Survey Path Coverage}

\begin{figure*}[bth]
%\epsscale{1.27}
\epsscale{1.2}
\plotone{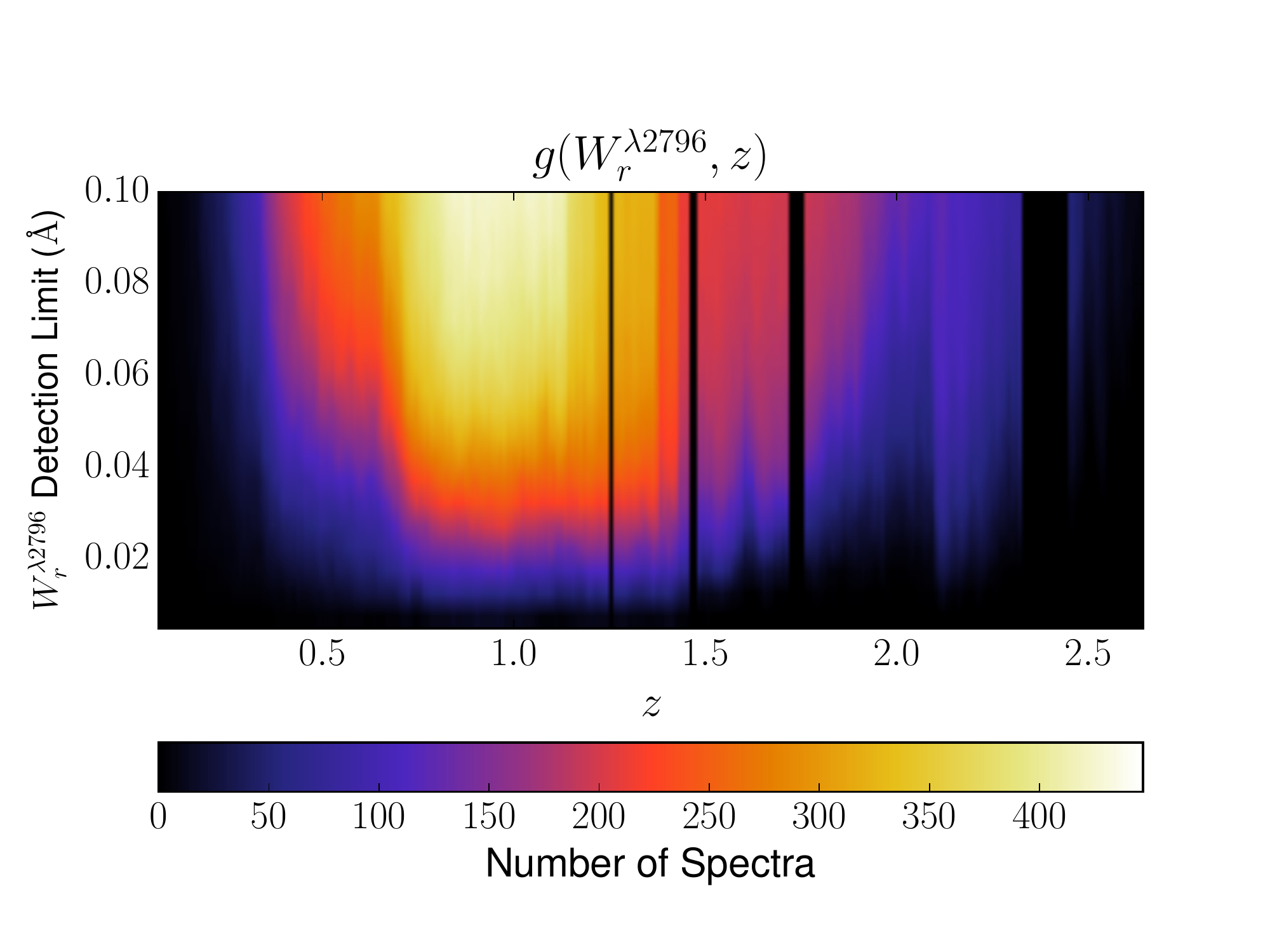}
\caption{The function $g(W_r^{\lambda2796}, z)$ shown as a heat map with the colors representing the value of $g(W_r^{2796}, z)$. This is the number of spectra in which an absorption line of a given equivalent width and a given redshift may be detected according to the detection limit of the spectrum. The vertical black bars representing no redshift path length coverage show the omitted wavelength regions of the survey based upon contaminating telluric absorption features.}
\label{fig:gwz}
\end{figure*}

To evaluate the completeness and calculate the redshift path coverage of our survey, we use the derived $5\sigma$ equivalent width detection limit described at the end of Section~\ref{sec:detection} to determine the number of spectra in which we could detect an absorber of a given equivalent width and redshift. Figure~\ref{fig:gwz} shows the function $g(W_r^{\lambda2796}, z)$, defined as

\begin{equation}
g(W_j,z_k) = \sum_n H(z_k - z_n^{\mathrm{min}}) H(z_n^{\mathrm{max}} - z_k) H[W_j - 5\sigma_k / (1 + z_k)],
\label{eqn:gwz}
\end{equation}

\noindent where $H$ is the Heaviside step function, $z_n^{\mathrm{min}}$ and $z_n^{\mathrm{max}}$ are the minimum and maximum redshifts observed for the $n$th quasar spectrum, and where the sum extends over all quasar spectra in the sample. This heat map details the number of spectra in which a {\MgIIdblt} doublet could be detected as a function of the equivalent width detection limit and redshift. The vertical stripes with no redshift path coverage represent the omitted telluric absorption regions for our survey. The integral along a given $W_r^{2796}$ slice gives the total redshift path length available for the sample ($\Delta Z$).

% ================ dN\!/dz + dN\!/dX ================
\subsection{$dN\!/dz$ and $dN\!/dX$}
\label{dndzdndx}

The largest sample of quasar spectra originates from the Sloan Digital Sky Survey (SDSS), with more than $10^5$ spectra at present, which employs a spectrograph with an instrumental resolution around $69$~{\kms}, limiting SDSS absorption surveys to strong absorbers, with $W_r^{\lambda2796} \ge 0.3$~{\AA} \citep{Nestor2005,Zhu2013}. Conversely, previous studies of weak absorbers used small samples of quasar spectra, never exceeding 100 quasar spectra \citep{Steidel1992,Narayanan2007,Kacprzak2011}. In this paper, we aim to characterize the evolution of the incidence rate, number of absorbers per redshift path length, comoving line density, and cosmic mass density of all ${\MgII}$ absorbers from redshifts $0.18 < z < 2.57$.

The number of ${\MgII}$ absorbers per redshift path length and its associated variance are defined as

\begin{equation}
\frac{d N}{d z} = \sum_{i}\frac{1}{\Delta Z_i\,(W_r)},\quad \sigma^2_{\frac{d N}{d z}} = \sum_{i} \Big[\frac{1}{\Delta Z_i\,(W_r)}\Big]^2,
\label{eqn:dndz}
\end{equation}

\noindent
where we count the number of ${\MgII}$ absorbers, dividing by the total searched redshift path length ($\Delta Z$), defined as

\begin{equation}
\Delta Z_i\,(W_r) = \int_{z_1}^{z_2} g_i\,(W_r, z)\,dz,
\label{eqn:deltaz}
\end{equation}

\noindent
where $g_i(W_r, z)$ is the equivalent width sensitivity function at a given equivalent width detection limit shown in Equation~\ref{eqn:gwz}. The function $g(W_r, z)$, first formulated in \cite{Lanzetta1987}, details the number of spectra in which an absorption feature with a given equivalent width may be detected at the $5\sigma$ level in a given redshift interval.

The comoving ${\MgII}$ line density and its associated variance are defined as

\begin{equation}
\frac{d N}{d X} = \sum_{i}\frac{1}{\Delta X_i\,(W_r)},\quad \sigma^2_{\frac{d N}{d X}} = \sum_{i} \Big[\frac{1}{\Delta X_i\,(W_r)}\Big]^2,
\label{eqn:dndx}
\end{equation}

\noindent
where we count the number of ${\MgII}$ absorbers, dividing by the total searched absorption path ($\Delta X$), defined as

\begin{equation}
\Delta X_i\,(W_r) = \int_{z_1}^{z_2} g_i\,(W_r, z) \frac{(1 + z)^2}{\sqrt{\Omega_M (1 + z)^3 + \Omega_{\Lambda}}}\,dz,
\label{eqn:deltax}
\end{equation}

\noindent
where $\Omega_M$ is the cosmic matter density, and $\Omega_{\Lambda}$ is the cosmic density attributed to dark energy. Counting with respect to $\Delta X$ accounts for both cosmological expansion along the line of sight and the transverse separation of objects with unchanging number density and cross section, allowing for more consistent comparisons across redshift.

% MURPHY has trouble understanding the last sentence.

\begin{figure*}[bth]
\epsscale{1.17}
\plottwo{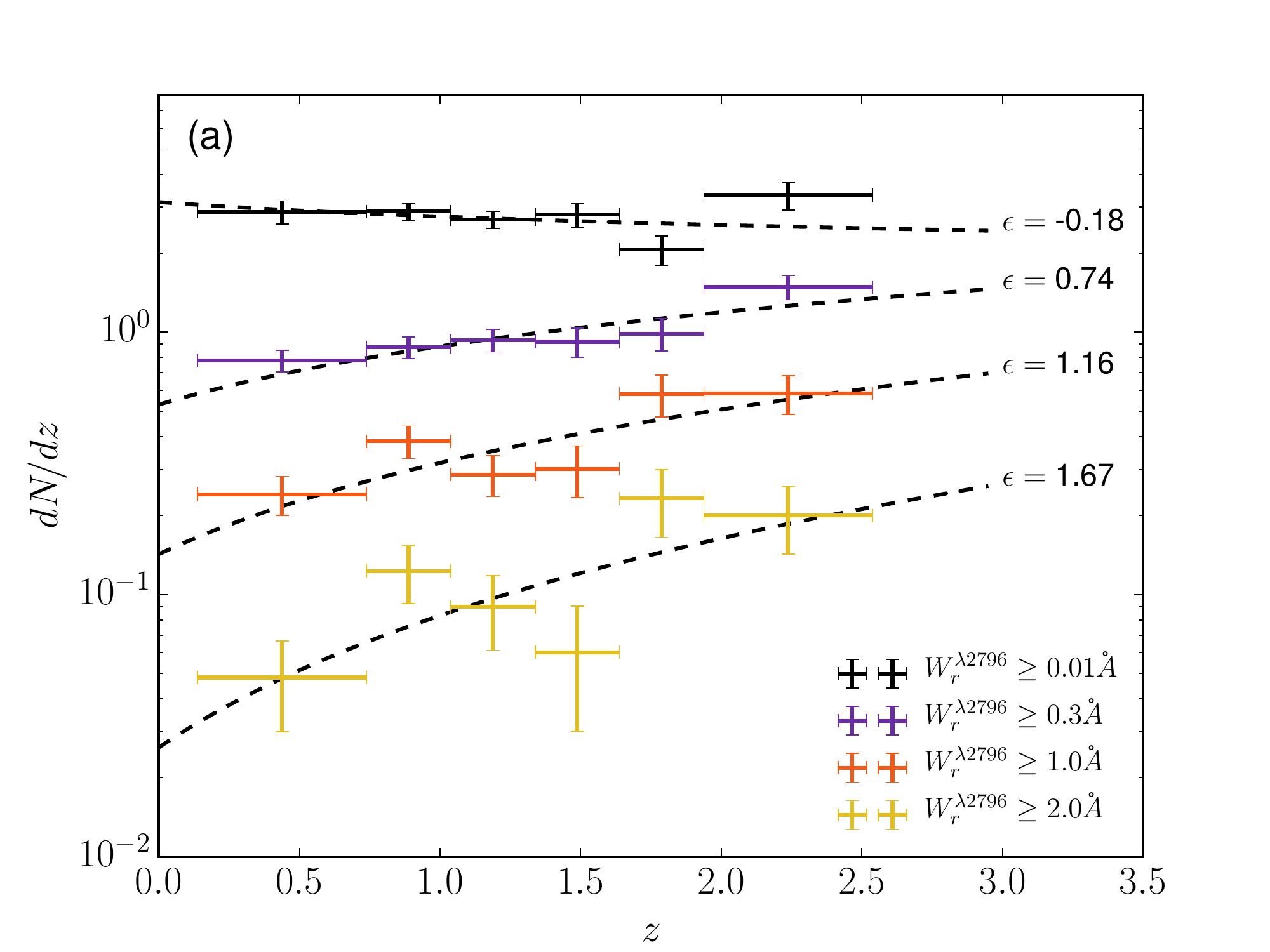}{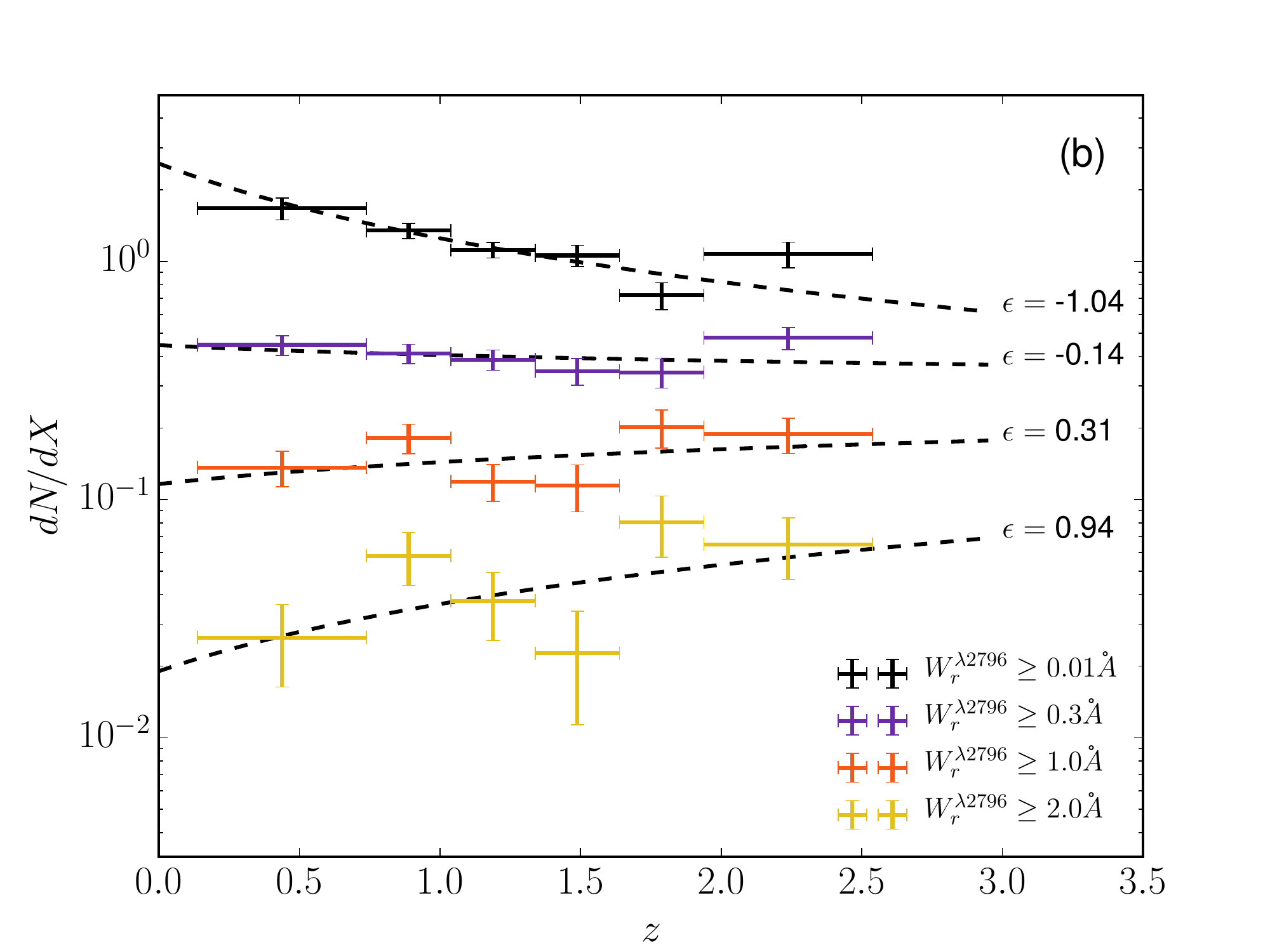}
\caption{(a) $dN\!/dz$ and (b) $dN\!/dX$ as a function of redshift for different minimum equivalent width thresholds, $W_{r,\mathrm{min}}^{\lambda2796}$. Colors represent different $W_{r,\mathrm{min}}^{\lambda2796}$. The black dotted lines are fits to the distribution of the functional form $f(z) = \frac{c}{H_o} n_0\,\sigma_0\,(1 + z)^{\epsilon}$, with the best fit $\epsilon$ value labelled. We see increasing values of $\epsilon$ with increasing equivalent width, driven by an enhancement of stronger ${\MgII}$ absorbers around redshift 2 compared to lower redshifts. Vertical error bars represent $1\sigma$ uncertainties in each bin.}
\label{fig:dndz_dndx}
\end{figure*}

In Figure~\ref{fig:dndz_dndx}, we plot $dN\!/dz$ and $dN\!/dX$, respectively, as a function of redshift for different minimum equivalent width thresholds, such that detected ${\MgII}$ absorbers have equivalent widths greater than $W_{r,\mathrm{min}}^{\lambda2796}$. Error bars in each bin represent $1\sigma$ uncertainties calculated according to Equations~\ref{eqn:dndz} and~\ref{eqn:dndx}. Dotted lines are fit according to the analytical form which allows for redshift evolution in $dN\!/dX$, defined as,

\begin{equation}
f(z) \equiv  \frac{c}{H_o} n(z)\,\sigma(z) =  \frac{c}{H_o} n_0\,\sigma_0\,(1+z)^{\epsilon},
\label{eqn:dndxfit}
\end{equation}

\noindent
where $c$ is the speed of light, $H_o$ is the Hubble Constant, $n_0$ is the comoving number density of ${\MgII}$ absorbers at $z = 0$, $\sigma_0$ is the absorbing cross-section at $z = 0$, and $\epsilon$ is the evolution parameter, defined as the power dependence of $dN\!/dX$ on redshift. The product $n_0\,\sigma_0$ represents a comoving opacity of ${\MgII}$-selected absorption line systems by virtue of the units, which are an inverse path length, and the analogous absorption coefficient to describe the opacity of material in stellar atmospheres. The full product of $\frac{c}{H_o} n_0\,\sigma_0$ then represents the comoving Hubble optical depth for ${\MgII}$ absorbers. We find that the best-fit value of $\epsilon$ is negative when analyzing the full sample of ${\MgII}$ absorbers, including all detections with measured equivalent widths above $W_r^{\lambda2796} > 0.01$~{\AA}. The evolution parameter, $\epsilon$, then increases with subsequently larger minimum equivalent width thresholds, becoming positive for absorbers with $W_{r,\mathrm{min}}^{\lambda2796} > 1.0$~{\AA}. This trend is driven primarily by an enhancement in $dN\!/dX$ for the strongest ${\MgII}$ absorbers around $z \sim 2$, relative to lower redshifts. Conversely, at low redshift we observe more weak ${\MgII}$ absorbers per absorption path length. We show in Table~\ref{tab:fitparams} the values of the fit parameters for varying $W_{r,\mathrm{min}}^{\lambda2796}$, along with their $1\sigma$ uncertainties.

%, $\tau_H({\MgII})$

\begin{deluxetable}{ccc}
\tablecolumns{3}
\tablewidth{0pt}
\tablecaption{Parameterization of $dN\!/dX$ \label{tab:fitparams}}
\tablehead{
  %\colhead{(1)} &
  %\colhead{(2)} &
  %\colhead{(3)} \\[2pt]
  \colhead{$W_{r,\mathrm{min}}^{2796}$} &
  \colhead{$\frac{c}{H_o}n_0\,\sigma_0$} &
  \colhead{$\epsilon$} \\[1pt]
  \colhead{[{\AA}]} &
  \colhead{} &
  \colhead{} }

\startdata
0.01 & 2.583 $\pm$ 0.827 & -1.04 $\pm$ 0.38 \\[3pt]
0.30 & 0.446 $\pm$ 0.076 & -0.14 $\pm$ 0.21 \\[3pt]
1.00 & 0.116 $\pm$ 0.043 & 0.31 $\pm$ 0.44 \\[3pt]
0.01 & 0.019 $\pm$ 0.014 & 0.94 $\pm$ 0.85
\enddata
\end{deluxetable}

\begin{figure*}[bth]
\epsscale{1.17}
\plottwo{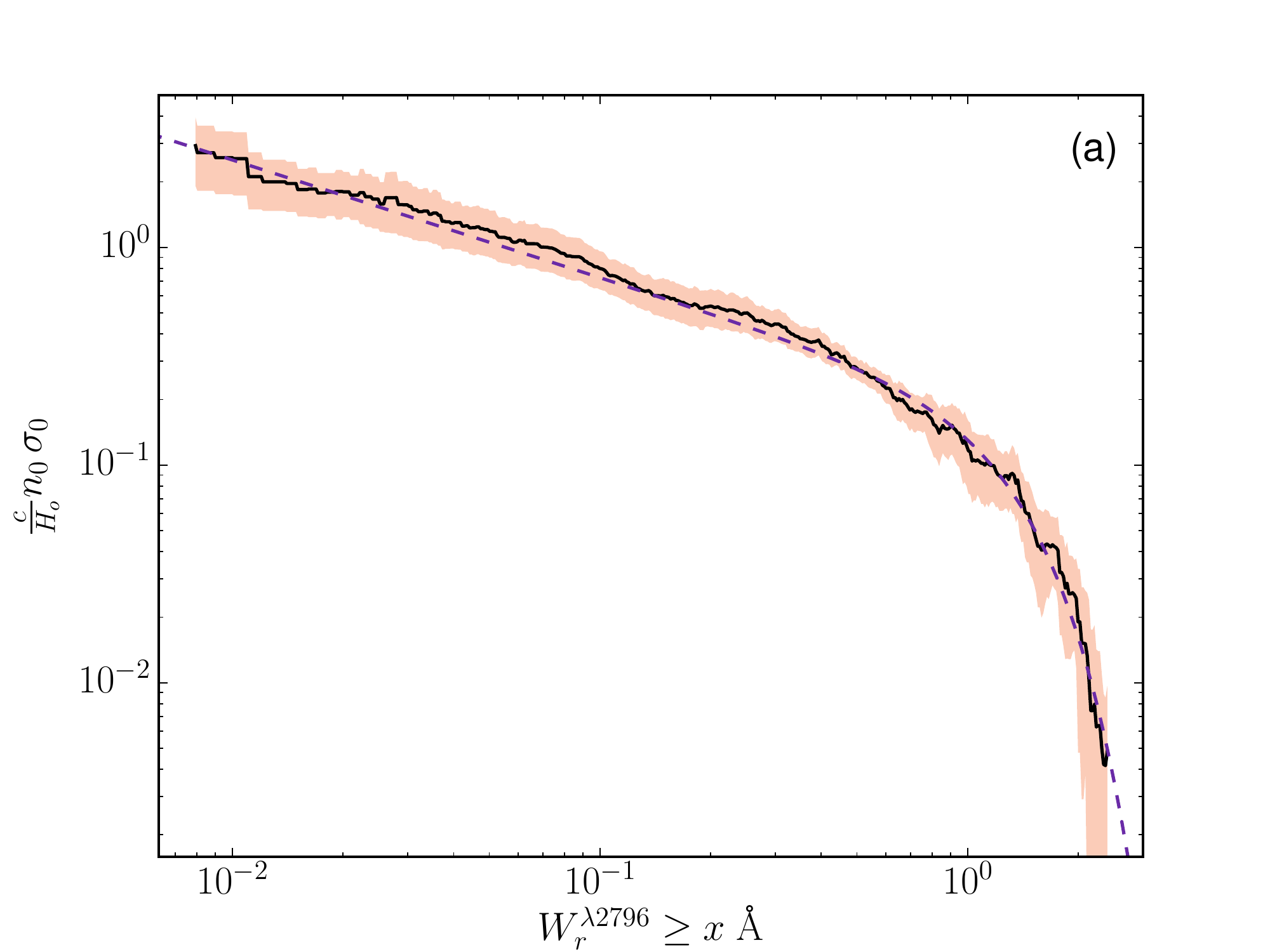}{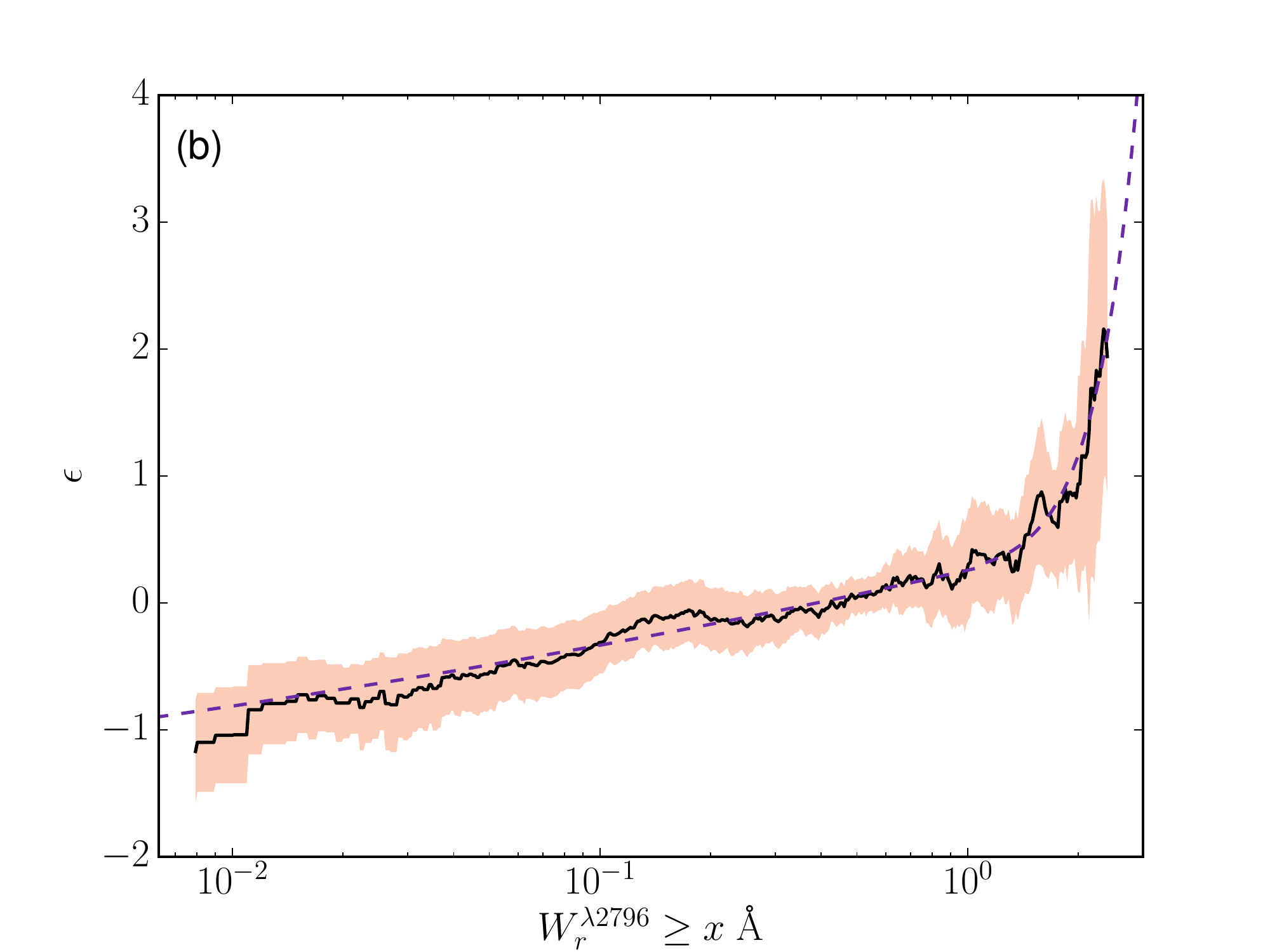}
\caption{(a) The comoving number density of absorbers multiplied by the absorbing cross-section, derived by fitting Equation~\ref{eqn:dndxfit} to $dN\!/dX$, as a function of $W_{r,\mathrm{min}}^{\lambda2796}$ with shaded $1\sigma$ uncertainties. As we examine samples with increasing minimum ${\MgII}$ equivalent width thresholds, either the space density of absorbing cloud structures decreases, the absorbing cross-section decreases, or both parameters decrease. (b) The redshift evolution parameter, $\epsilon$, as a function of $W_{r,\mathrm{min}}^{\lambda2796}$. Weak ${\MgII}$ absorbers are more abundant at low redshift, leading to a negative coefficient $\epsilon$. Absorbers with equivalent widths near $0.3$~{\AA} do not evolve, with $\epsilon \simeq 0$. Strong ${\MgII}$ absorbers evolve away at low redshift, showing a large positive $\epsilon$ increasing towards $z \sim 2$.}
\label{fig:nsigmaepsilon}
\end{figure*}

In Figure~\ref{fig:nsigmaepsilon}, we show the values of $\frac{c}{H_o}n_0\,\sigma_0$ and $\epsilon$ as a function of $W_{r,\mathrm{min}}^{\lambda2796}$. The shaded red areas represent the $1\sigma$ standard deviations derived from the fits to the $dN\!/dX$ distribution. We show first that the comoving Hubble optical depth of ${\MgII}$ absorbers decreases as a function of $W_r^{\lambda2796}$. This implies that, per unit absorption path length, there are fewer high equivalent width ${\MgII}$ absorbers, and/or that they exist in smaller absorbing structures. We also show that the slope of the redshift dependence, $\epsilon$, increases as a function of increasing $W_{r,\mathrm{min}}^{\lambda2796}$. This evolution parameter, $\epsilon$, changes from negative to positive toward higher equivalent width ${\MgII}$ absorbers, implying that strong ${\MgII}$ absorbers evolve away, decreasing in relative number per absorption path length, from $z = 2$ to present. Conversely, weak ${\MgII}$ absorbers build up over time, increasing in relative number per absorption path length from $z = 2$ to present. We observe no evolution with redshift in absorbers with equivalent widths between $0.3 < W_r^{\lambda2796} < 1$~{\AA}.

We provide a parameterized fit to $\frac{c}{H_o}\,n_0\,\sigma_0$ and $\epsilon$ as a function of $W_{r,\mathrm{min}}^{\lambda2796}$. In Figure~\ref{fig:nsigmaepsilon}(a), we adopt a power-law with a generalized exponential decay to model the $\frac{c}{H_o}\,n_0\,\sigma_0$ distribution, defined as,

\begin{equation}
\frac{c}{H_o}n_0\,\sigma_0\,(\,\psi\,) = \Psi^* (\,\psi\,)^{\alpha} e^{-\,\psi\,^{\beta}} ,
\label{eqn:nsigmafit}
\end{equation}

\noindent where $\psi = W_{r,\mathrm{min}}^{\lambda2796} / W_{r,\mathrm{min}}^*$ to simplify the equation. The best fit parameters are $\Psi^* = 0.24 \pm 0.01$, $W_{r,\mathrm{min}}^* = 1.19 \pm 0.02$, $\alpha = -0.49 \pm 0.01$, and $\beta = 1.50 \pm 0.05$. This parameterization resembles a Schechter function, but we required an exponential drop-off at the high end faster than $e^{-x}$, which manifests itself in the form of $\beta$. Next, in Figure~\ref{fig:nsigmaepsilon}(b), we fit a broken power-law to the $\epsilon$ distribution, defined as,

\begin{equation}
%\[
\epsilon\,(W_{r,\mathrm{min}}^{\lambda2796}) =
\left\{
\!
\begin{aligned}
%3.64\,(W_{r,\mathrm{min}}^{\lambda2796})^{0.08}\, -\, 3.44 &\,\, \text{ if }\,\, W_{r,\mathrm{min}}^{\lambda2796} < 1.1~{\angstrom} \\[2mm]
%0.07\,(W_{r,\mathrm{min}}^{\lambda2796})^{3.72}\, +\, 0.15 &\,\, \text{ if }\,\, W_{r,\mathrm{min}}^{\lambda2796} \ge 1.1~{\angstrom},
a_1\,(\,W_{r,\mathrm{min}}^{\lambda2796}\,)^{\gamma_1} + b_1 &\,\, \text{ if }\,\, W_{r,\mathrm{min}}^{\lambda2796} < 1.1~{\angstrom} \\[2mm]
a_2\,(\,W_{r,\mathrm{min}}^{\lambda2796}\,)^{\gamma_2} + b_2 &\,\, \text{ if }\,\, W_{r,\mathrm{min}}^{\lambda2796} \ge 1.1~{\angstrom},
\label{eqn:epsilonfit}
\end{aligned}
\right.
%\]
\end{equation}

\noindent where the fit parameters for minimum equivalent width thresholds below $1.1$~{\AA} are $a_1 = 3.13 \pm 1.49$, $\gamma_1 = 0.09 \pm 0.05$, and $b_1 = 2.87 \pm 1.49$. The fit parameters for the power-law with $W_{r,\mathrm{min}}^{\lambda2796} \ge 1.1$~{\AA} are $a_2 = 0.07 \pm 0.01$, $\gamma_2 = 3.73 \pm 0.25$, and $b_2 = -0.21 \pm 0.03$. Combining the fits to $\frac{c}{H_o}\,n_0\,\sigma_0$ and $\epsilon$, we now have an analytic parameterization of $dN\!/dX$ as a function of $W_{r,\mathrm{min}}^{\lambda2796}$, and $z$, of the form,

\begin{equation}
\frac{dN}{dX}(W_{r,\mathrm{min}}^{\lambda2796}, z) = \Psi^* \psi^{\alpha} e^{-\psi^{\beta}} (1 + z)^{\epsilon(W_{r,\mathrm{min}}^{\lambda2796})}.
\label{eqn:dndxanalytic}
\end{equation}

\noindent This function can be used in future semi-analytic models to parameterize the physical properties of ${\MgII}$ absorbers in galaxy halos (e.g.~\cite{Shattow2015}).

% ================ EW's + logN ================
\subsection{Equivalent Width Frequency Distribution}
\label{sec:ewdistro}

To calculate the equivalent width frequency distribution $f(W)$, the number of absorbers of a given equivalent width per unit path density, we calculate $dN\!/dz$ or $dN\!/dX$ for each equivalent width bin and divide by the bin width. We split the sample into four redshift regimes, ensuring that the number of absorbers in each redshift subsample remains constant. The result is a characteristic number of ${\MgII}$ absorbers per redshift or absorption path length per equivalent width.

\begin{figure*}[bth]
\epsscale{1.17}
\plottwo{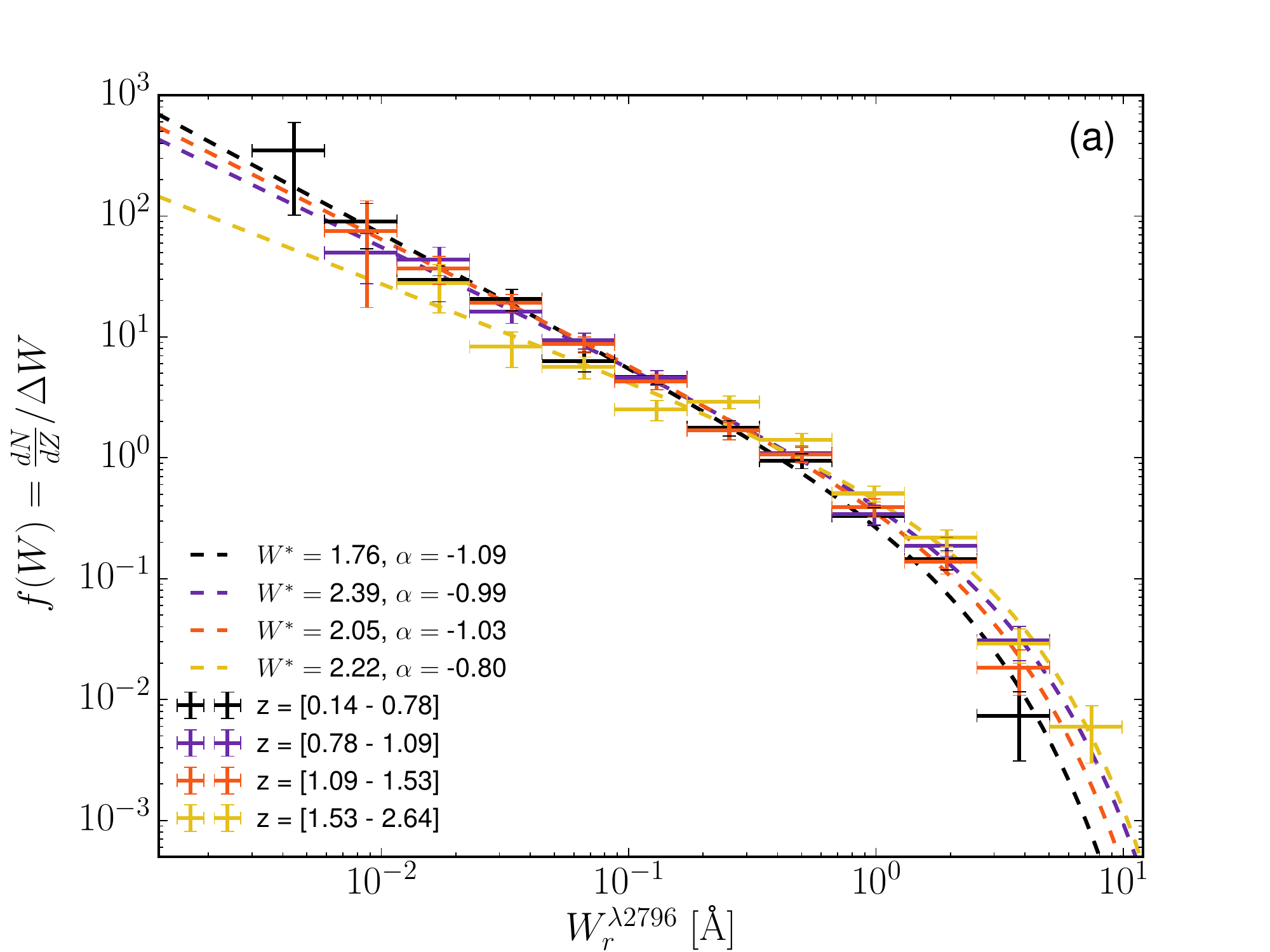}{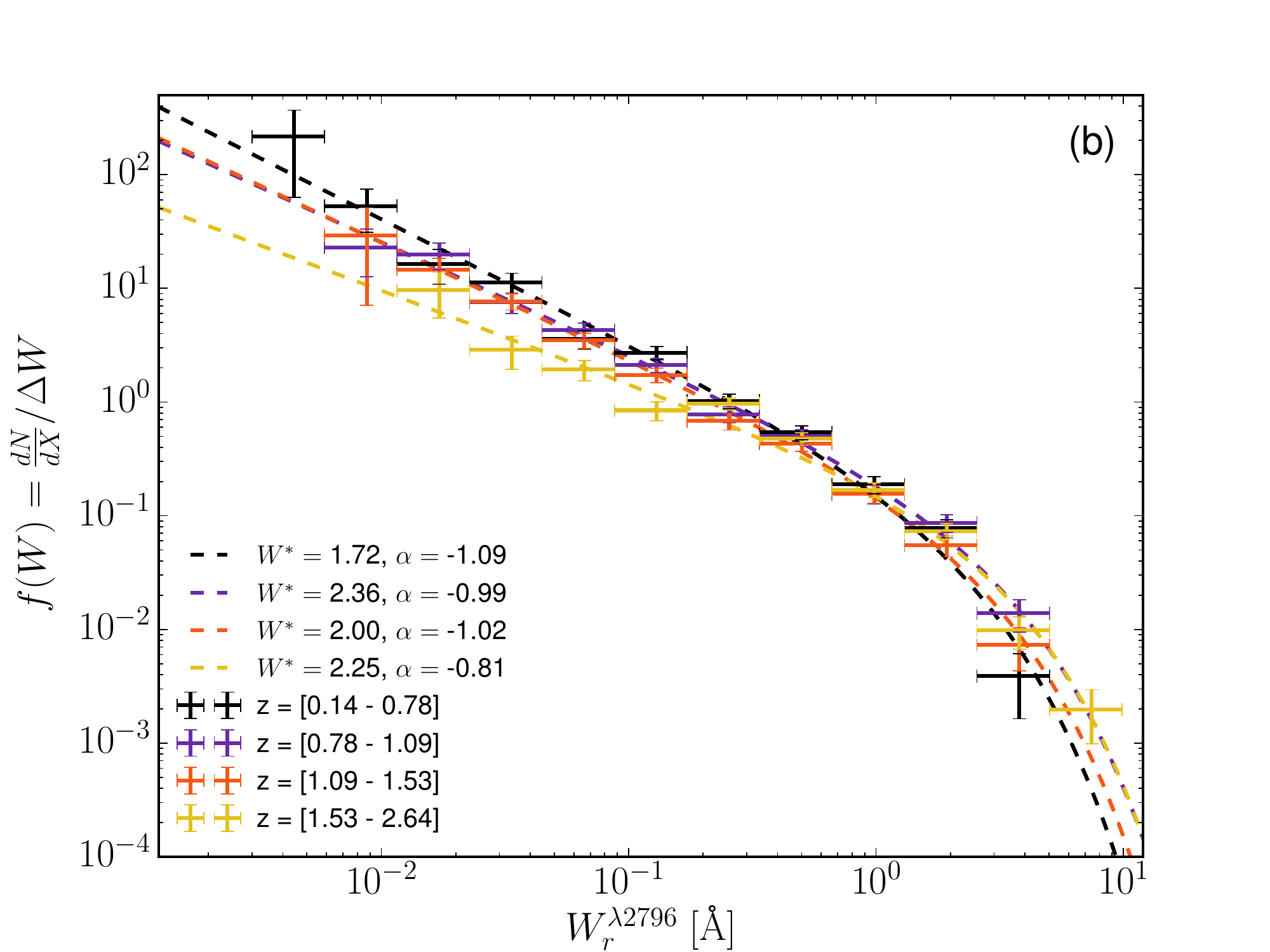}
\caption{(a) The equivalent width distribution of ${\MgII}$ absorbers, defined as the redshift path density ($dN\!/dz$) in each equivalent width bin divided by the bin width. (b) The equivalent width distribution, defined as the comoving line density ($dN\!/dX$) in each equivalent width bin divided by the bin width. Error bars represent $1\sigma$ uncertainties in each bin. We fit each distribution with a Schechter function, capturing the self-similar power-law behavior of weak ${\MgII}$ absorbers and the exponential power-law cutoff when observing the strongest ${\MgII}$ systems.}
\label{fig:ewdistro}
\end{figure*}

In Figure~\ref{fig:ewdistro}, we plot the equivalent width frequency distribution with respect to either $dN\!/dz$ or $dN\!/dX$. We fit each distribution with a Schechter function of the form,

\begin{equation}
\Phi (W_r) = \Phi^* \left(\frac{W_r}{W_r^*}\right)^{\alpha} e^{-W_r / W_r^*} ,
\label{eqn:schechter}
\end{equation}

\noindent where $\Phi^*$ is the normalization, $\alpha$ is the low equivalent width power-law slope, and $W_r^*$ is the turnover point in the distribution where the low equivalent width power-law slope transitions into an exponential cutoff. Table~\ref{tab:schechterew} shows the values of $\Phi^*$, $W_r^*$, and $\alpha$, along with their associated $1\sigma$ uncertainties derived from the fitting routine. This functional fit is motivated by papers such as \cite{Kacprzak2011MgII}, where the authors seek to combine previous surveys of strong ${\MgII}$ absorbers, in which exponential fits were preferred, and surveys of weak ${\MgII}$ absorbers, where power-laws best fit the equivalent width distribution. The power-law nature of the distribution of weak absorbers in our survey is apparent, and the exponential cutoff is motivated by physical limits to the size, density, and velocity widths of ${\MgII}$ absorbing clouds. Examining the distribution as a function of redshift, we find the low equivalent width slope becomes more shallow at $z \sim 2$ compared with the present epoch, with $\alpha = -1.09$ in our subsample with $0.14 \le z < 0.78$ and $\alpha = -0.81$ in our subsample with $1.53 < z \le 2.64$. We observe fewer weak ${\MgII}$ absorbers and more strong ${\MgII}$ absorbers per redshift/comoving absorption path length at $z \sim 2$ than we do at $z \sim 0.5$.

% As shown in Table 2, alpha is approximately -1 and then it changes drastically at z = 2 with $\alpha \sim 0.8$

\begin{deluxetable}{lccc}
\tablecolumns{4}
\tablewidth{0pt}
\tablecaption{Schechter Fit to $f(W) = \frac{dN}{dX} / \Delta W$ \label{tab:schechterew}}
\tablehead{
  %\colhead{(1)} &
  %\colhead{(2)} &
  %\colhead{(3)} \\[2pt]
  \colhead{Redshift Range} &
  \colhead{$\Phi^*$} &
  \colhead{$W^*$} &
  \colhead{$\alpha$} \\[1pt]
  \colhead{} &
  \colhead{} &
  \colhead{$[${\AA}$]$} &
  \colhead{}
}
\startdata
$0.14 - 0.78$  & $0.15 \pm 0.10$ & $1.72 \pm 0.68$ & $-1.09 \pm 0.09$ \\[3pt]
$0.78 - 1.09$  & $0.12 \pm 0.06$ & $2.36 \pm 0.81$ & $-0.99 \pm 0.06$ \\[3pt]
$1.09 - 1.53$  & $0.11 \pm 0.03$ & $2.00 \pm 0.35$ & $-1.02 \pm 0.04$ \\[3pt]
$1.53 - 2.64$  & $0.12 \pm 0.08$ & $2.25 \pm 0.87$ & $-0.81 \pm 0.12$
\enddata
\end{deluxetable}

\subsection{Column Density Distribution}
\label{sec:logndistro}

To calculate the column density distribution, the number of absorbers of a given column density per unit path density, we calculate $dN\!/dz$ or $dN\!/dX$ for each column density bin and divide by the bin width. The result is a characteristic number density of ${\MgII}$ absorbers per redshift or absorption path length as a function of their column densities. It should be noted that at high column densities near $\log (N({\MgII})) = 15~\mathrm{cm^{-2}}$, the measured column densities are lower limits as the AOD method cannot constrain the true column when the absorption line becomes saturated.

\begin{figure*}[bth]
\epsscale{1.17}
\plottwo{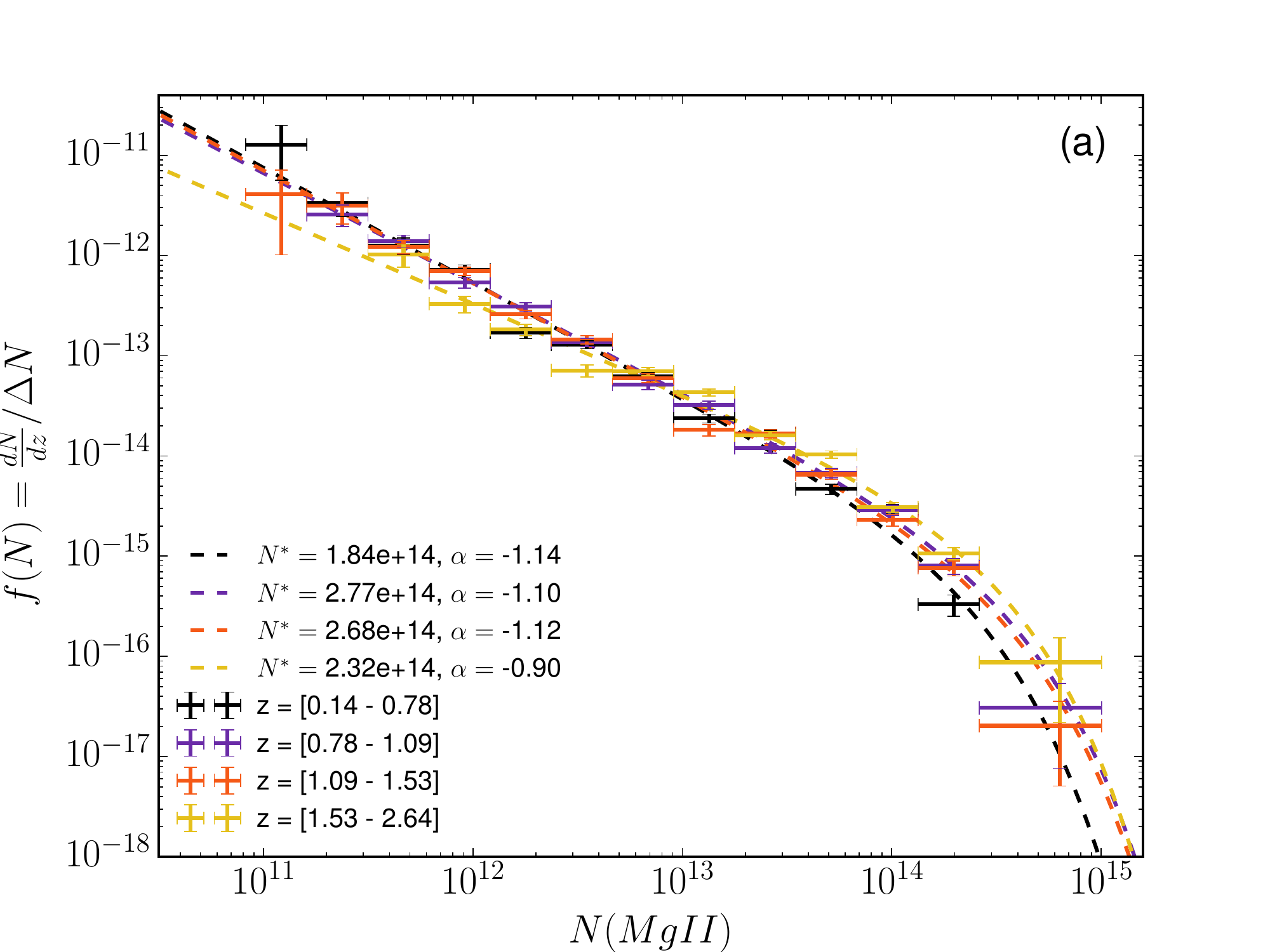}{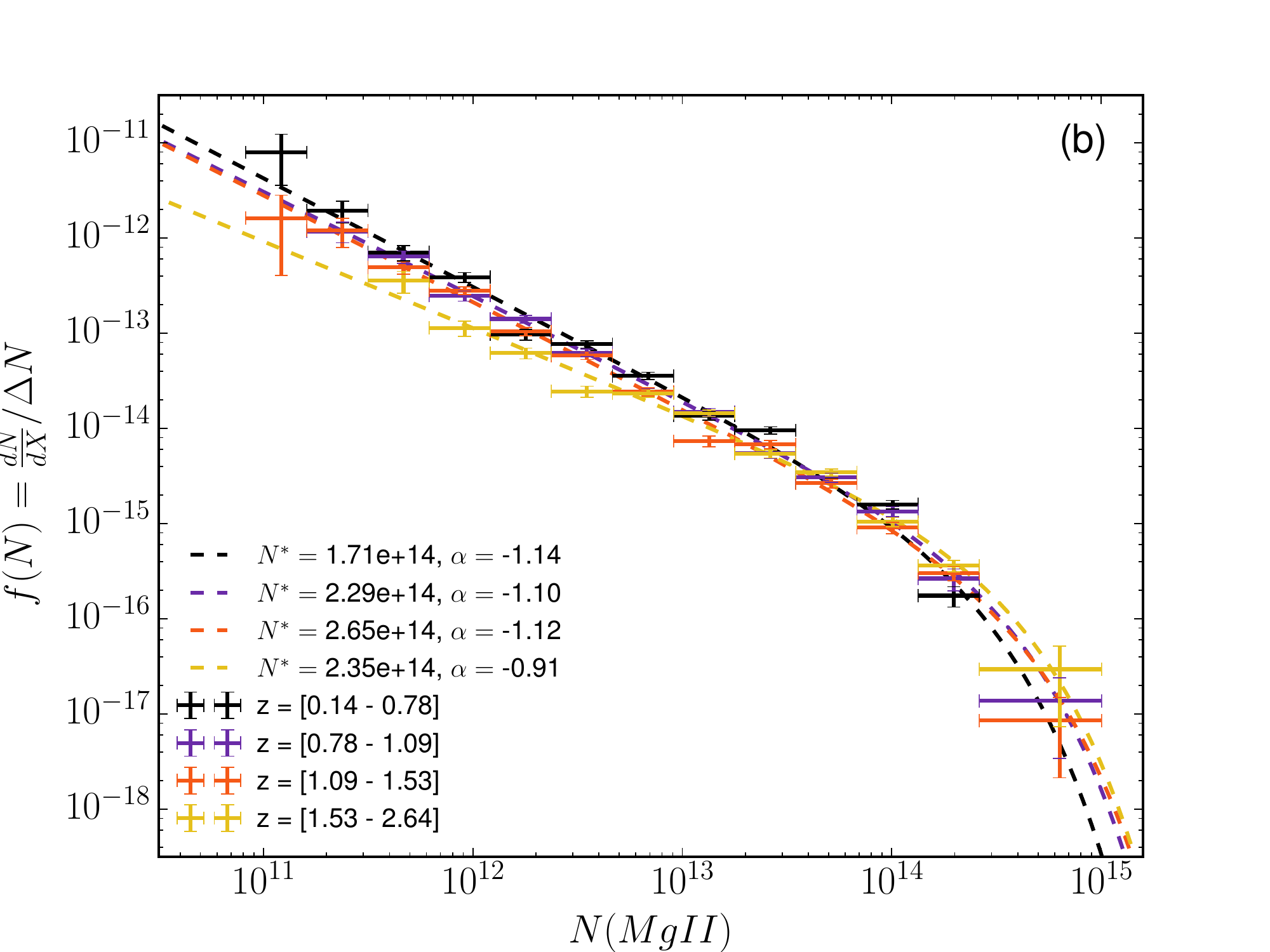}
\caption{(a) The column density distribution of ${\MgII}$ absorbers, defined as the redshift path density ($dN\!/dz$) in each column density bin dividided by the bin width. (b) The comoving line density ($dN\!/dX$) in each column density bin divided by the bin width. We fit this distribution with a Schechter function to accurately parameterize the low column density power-law slope and the exponential cutoff and high column densities.}
\label{fig:logndistro}
\end{figure*}

In Figure~\ref{fig:logndistro}, we plot the column density frequency distribution using either $dN\!/dz$ or $dN\!/dX$. Again, we fit this distribution with a Schechter function of the same form as Equation~\ref{eqn:schechter}, except with equivalent width replaced with column density. Table~\ref{tab:schechtern} shows the values of $\Phi^*$, $N^*$, and $\alpha$, along with their associated $1\sigma$ uncertainties. We find again that the low column density slope is shallower near $z \sim 2$ than at $z \sim 0.5$. Due to saturation effects, the highest column density measurements are lower limits; therefore, the final high column density bin in $f(N)$ should be regarded as an upper limit. These limits are taken into account in the functional fitting procedures.

\begin{deluxetable}{lccc}
\tablecolumns{4}
\tablewidth{0pt}
\tablecaption{Schechter Fit to $f(N) = \frac{dN}{dX} / \Delta N$ \label{tab:schechtern}}
\tablehead{
  %\colhead{(1)} &
  %\colhead{(2)} &
  %\colhead{(3)} \\[2pt]
  \colhead{Redshift Range} &
  \colhead{$\Phi^*$} &
  \colhead{$N^*$} &
  \colhead{$\alpha$} \\[1pt]
  \colhead{} &
  \colhead{$[\times 10^{-16}]$} &
  \colhead{$[\times 10^{14}~\mathrm{cm}^{-2}]$} &
  \colhead{}
}
\startdata
$0.14 - 0.78$ & $8.79 \pm 8.19$ & $1.71 \pm 0.97$ & $-1.14 \pm 0.08$ \\[3pt]
$0.78 - 1.09$ & $6.41 \pm 2.43$ & $2.29 \pm 0.56$ & $-1.10 \pm 0.03$ \\[3pt]
$1.09 - 1.53$ & $4.15 \pm 2.40$ & $2.65 \pm 1.00$ & $-1.12 \pm 0.04$ \\[3pt]
$1.53 - 2.64$ & $7.98 \pm 6.51$ & $2.35 \pm 1.32$ & $-0.91 \pm 0.08$
\enddata
\end{deluxetable}

% ================ Omega_MgII ================
\subsection{$\Omega_{\hbox{\scriptsize {\MgII}}}$}
\label{omegamgii}

We now aim to calculate the matter density of ${\MgII}$ absorbers across cosmic time. To do so, we employ the following customary equation relating the mass density of an ion as a fraction of the critical density today to the first moment of the column density distribution,

\begin{equation}
\Omega_{\hbox{\scriptsize {\MgII}}} = \frac{H_0\  m_{\hbox{\scriptsize {\MgII}}}}{c\ \rho_{c,0}} \int_{N_{min}}^{N_{max}}\, f (N_{\hbox{\scriptsize {\MgII}}})\, N_{\hbox{\scriptsize {\MgII}}}\, dN_{\hbox{\scriptsize {\MgII}}} ,
\label{eqn:omega}
\end{equation}

\noindent where $H_0$ is the Hubble constant today, $m_{\mathrm{Mg}} = 4.035 \times 10^{-23}~\mathrm{g}$, $c$ is the speed of light, $\rho_{c,0}$ is the critical density at present, $f(N_{\hbox{\scriptsize {\MgII}}})$ is the column density distribution of ${\MgII}$ absorbers, and $N_{\hbox{\scriptsize {\MgII}}}$ is the column density. Using our derived fit to the column density distribution, we are able to numerically integrate the first moment from $0 < \log N({\MgII}) < 20~\mathrm{cm^{-2}}$. The results are shown below in Figure~\ref{fig:omegamgii}. $1\sigma$ uncertainties are derived with a bootstrap Monte-Carlo method. We select random column densities, with replacement, from the sample of measured column densities for all of our ${\MgII}$ absorbers until we reach the sample size. We then recalculate the column density distribution, find the best parameterized Schechter fit, and then integrate and compute Equation~\ref{eqn:omega}. We perfom this task 1499 times to develop a statistical ensemble of values for $\Omega$, with this number of samples representing the underlying scatter in the $\Omega_{\hbox{\scriptsize {\MgII}}}$ distribution at the 99\% confidence level according to~\cite{Davidson2000bootstrap}. We take the standard deviation about the mean of this ensemble of simulated measurements as the $1\sigma$ uncertainty in $\Omega_{MgII}$. We find that the cosmic mass density of ${\MgII}$ increases from $\Omega_{\hbox{\scriptsize {\MgII}}} \simeq 0.8\times10^{-8}$ at $z \sim 0.5$ to $\Omega_{\hbox{\scriptsize {\MgII}}} \simeq 1.3\times10^{-8}$ at $z \sim 2$.

\begin{figure}[bth]
\epsscale{1.2}
\plotone{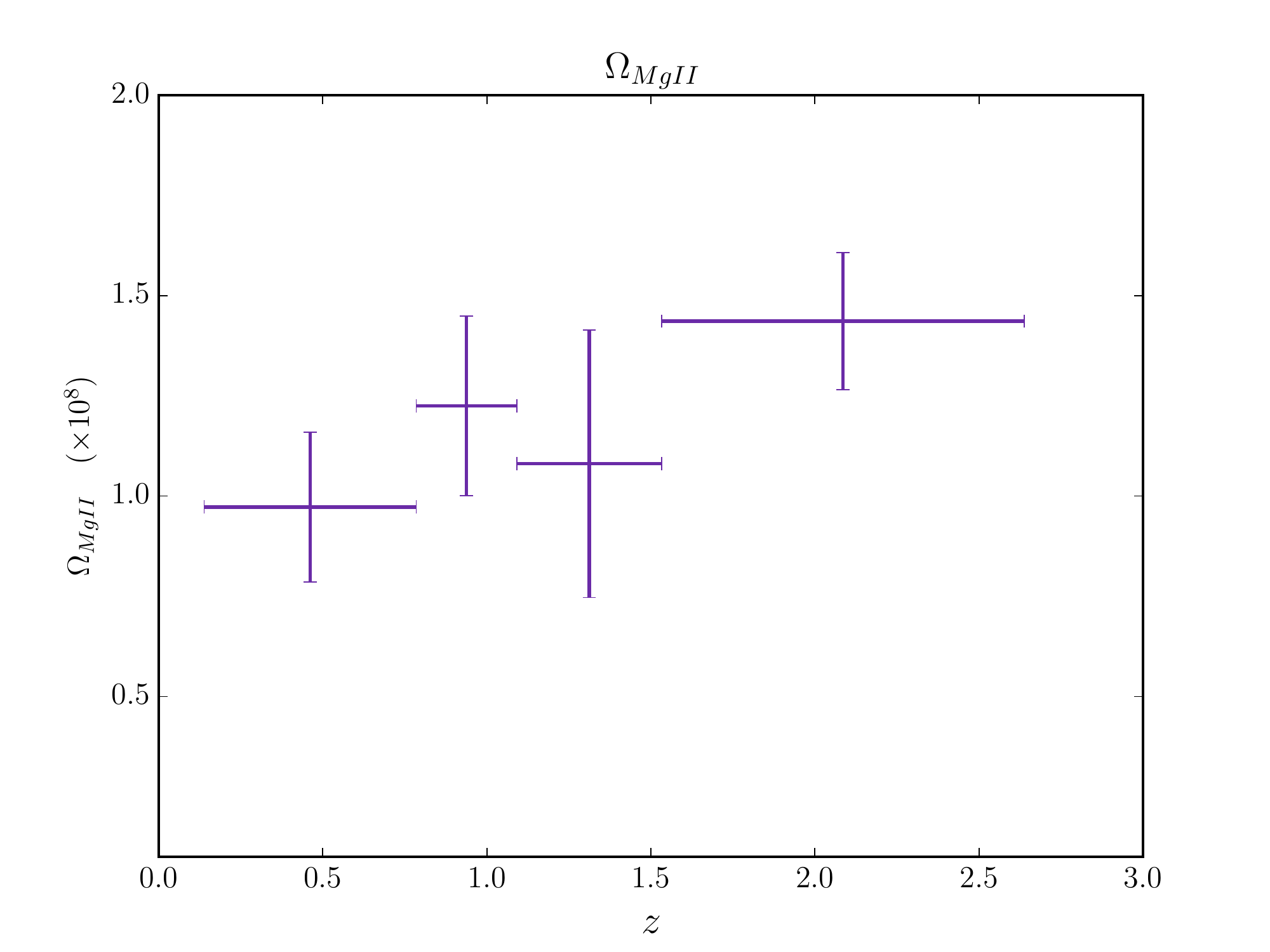}
\caption{$\Omega_{\hbox{\scriptsize {\MgII}}}$ as a function of redshift. The cosmic mass density of ${\MgII}$ stays roughly flat near a value of $1 \times 10^{-8}$, with a 0.5 dex increase from $z \sim 0.5$ to $z \sim 2$.}
\label{fig:omegamgii}
\end{figure}

% ================ DISCUSSION ================
\section{Discussion}
\label{sec:discussion}

We have shown a cosmic inventory of ${\MgII}$ absorbing gas from $0.1 < z < 2.6$, measuring $dN\!/dz$, $dN\!/dX$, the equivalent width distribution, and the column density distribution down to detection limits as low as $W_r^{\lambda2796} = 0.01$~{\AA}. We aim now to relate the properties of ${\MgII}$ absorbers and their evolution across cosmic time to other known evolutionary processes with the hope of gaining insight into the mechanisms which give rise to ${\MgII}$ absorbing gas.

\subsection{Evolution of ${\MgII}$ Distributions}

\cite{Narayanan2007} measured the evolution of weak ${\MgII}$ absorbers from $0.4 < z < 2.4$ in VLT/UVES spectra. They compared to \cite{Churchill1999}, who fitted the equivalent width frequency distribution with a power-law, and to \cite{Nestor2005}, who fitted an exponential to $f(W_r)$. In the case of weak absorbers at $z < 1.4$, \cite{Narayanan2007} found that a power-law with a slope of $\alpha = -1.04$ is a satisfactory fit, confirming the results of \cite{Churchill1999}. When they split their sample into low redshift, with detections between $0.4 < z < 1.4$, and high redshift, with detections between $1.4 < z < 2.4$, they found that the low redshfit sample remained consistent with a power-law but the high redshift sample was best fit by an exponential function. Our data show that a faint end power-law slope of $\alpha = -0.81$ is appropriate for the higher redshift subsample, and we note that this is also consistent with the data of~\cite{Narayanan2007}.

\begin{figure}[bth]
\epsscale{2.25}
\plottwo{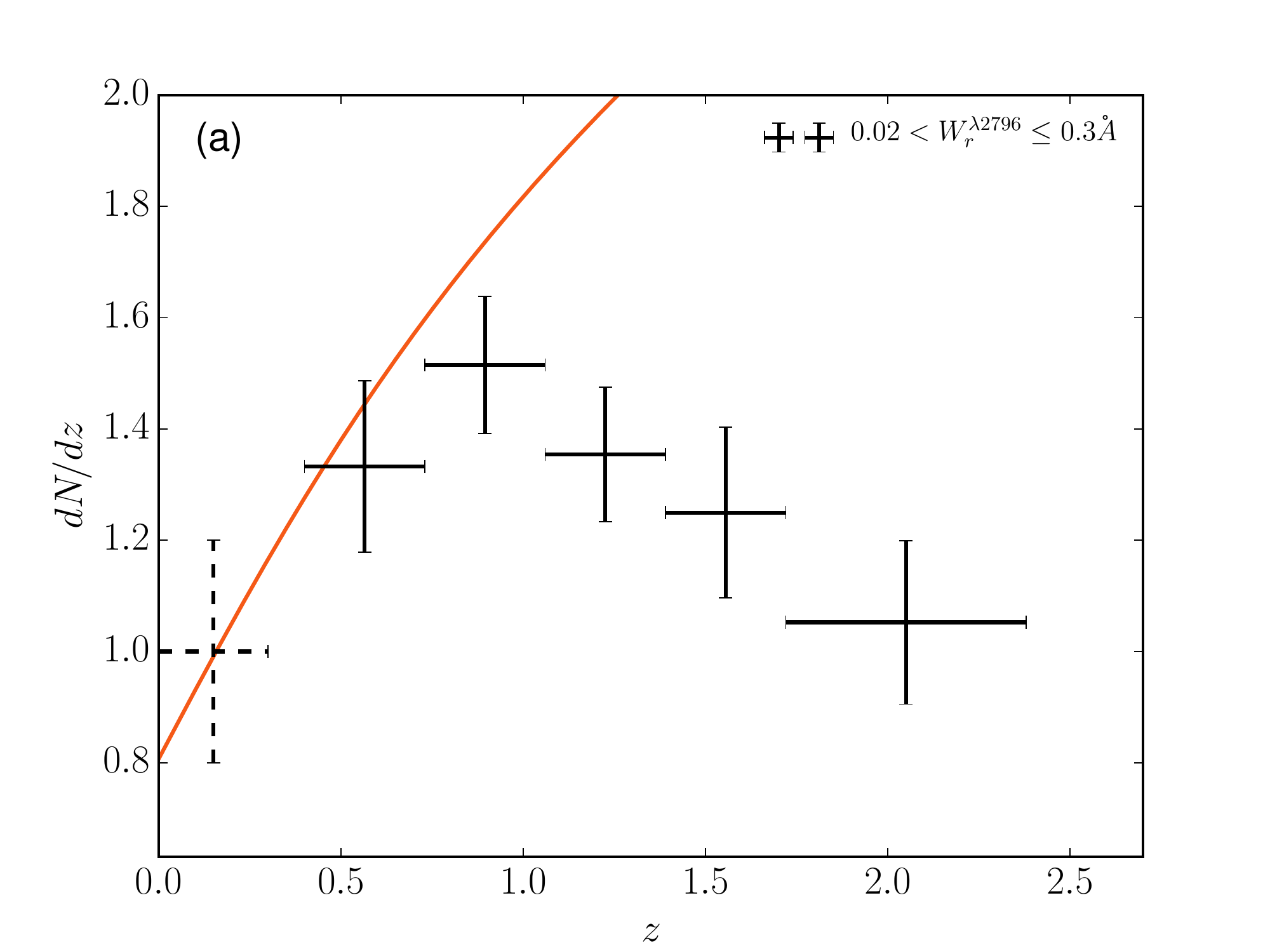}{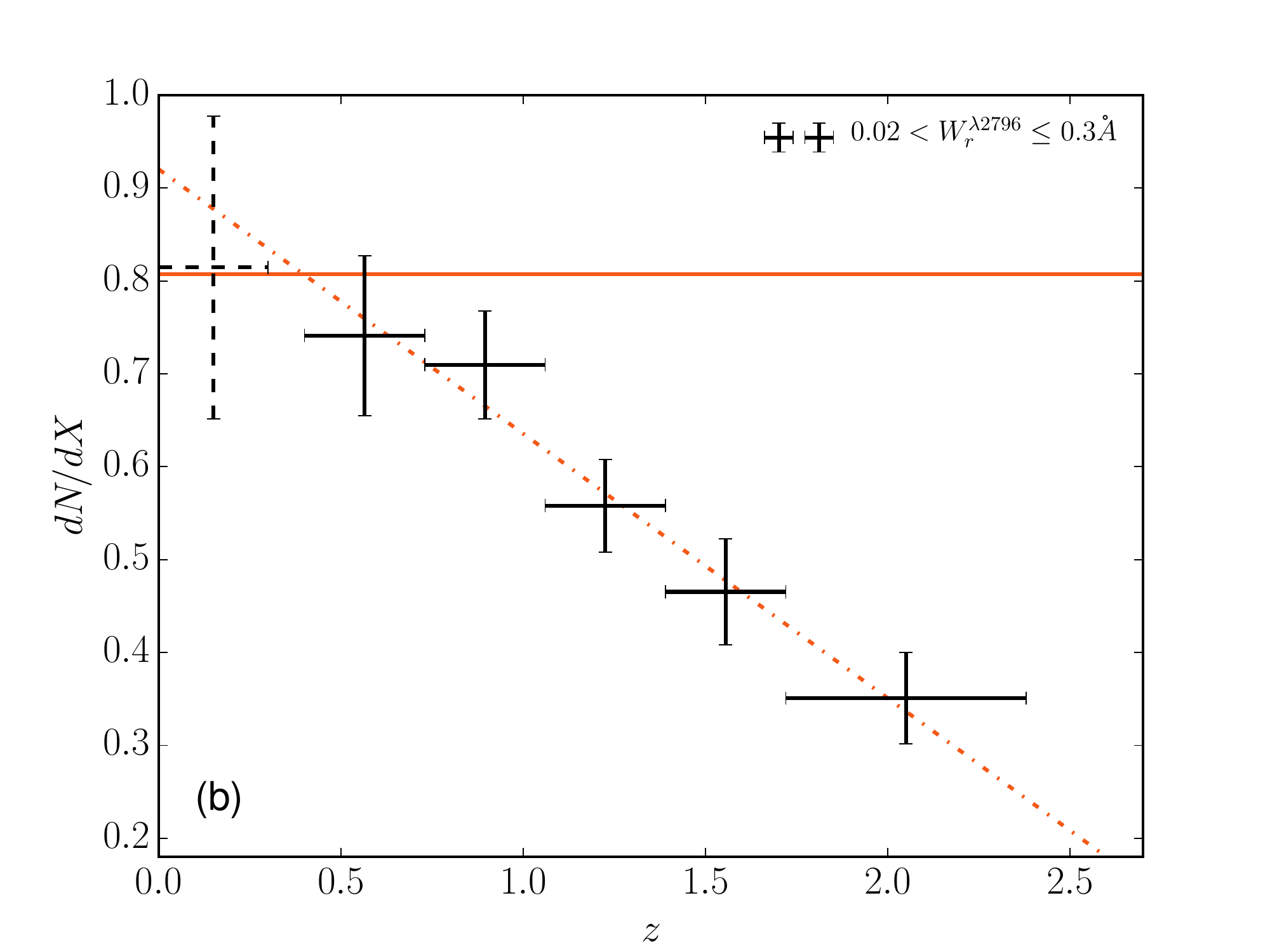}
\caption{(a) $dN\!/dz$ as a function of redshift for equivalent widths in the range $0.02 < W_{r}^{\lambda2796} \le 0.3$~{\AA}. We also include the survey data point from $0 < z < 0.3$ of~\cite{Narayanan2005} as shown by the dashed histogram. (b) $dN\!/dX$ as a function of redshift for the same population. Error bars represent $1\sigma$ uncertainties as calculated in Equations~\ref{eqn:dndz} and~\ref{eqn:dndx}. The red solid lines represent the no-evolution expectation, normalized at $z = 0.9$ and $dN\!/dz = 1.74$, matching~\cite{Narayanan2007}. The red dot-dashed line in panel (b) represents a linear fit to the binned data of the form $dN\!/dX = -0.28^{(\pm0.04)}z + 0.92^{(\pm0.04)}$. By this construction, $dN\!/dX = 0$ at $z = 3.3$, implying we should detect no weak ${\MgII}$ absorbers past this redshift. This equivalent width range comprises exclusively ``weak'' ${\MgII}$ absorbers.}
\label{fig:dndzbetween}
\end{figure}

\cite{Narayanan2007} also analyzed the evolution of $dN\!/dz$ with redshift for weak ${\MgII}$ absorbers. They found that the distribution follows the ``no evolution'' expectation; that is, the expected number density for a nonevolving population of absorbers in a $\mathrm{\Lambda CDM}$ universe, at redshifts less than $z = 1.5$. At higher redshift, they found that $dN\!/dz$ for weak absorbers decreases below the no evolution expectation. In Figure~\ref{fig:dndzbetween}(a) we make a direct comparison with \cite{Narayanan2007}, showing $dN\!/dz$ for $0.02 \le W_r^{\lambda2796} < 0.3$~{\AA} binned in the same manner as their Figure 4. Here, we observe that the apparent peak in $dN\!/dz$ for weak absorbers occurs near $z = 0.75$, as opposed to $z = 1.2$ in \cite{Narayanan2007}. However, the overall shape of $dN\!/dz$ for this low equivalent width population remains in good agreement, with the number of weak absorbers per redshift path length rising from $z = 0.4$ to $z \sim 1.0$, following the no evolution expectation, and then falling well below that expectation from $z \sim 1.0$ to $z \sim 2.4$. In Figure~\ref{fig:dndzbetween}(b), we show $dN\!/dX$ for weak ${\MgII}$ absorbers. Here, we clearly see evolution in the distribution of low equivalent width absorbers, showing that the comoving line density of weak ${\MgII}$ absorbers steadily decreases as a function of redshift from $z = 0.4$ to $z = 2.4$. The ``no evolution'' assumption would be a perfectly flat distribution, as shown by the solid red line. We also fit a linear approximation to the binned data of Figure~\ref{fig:dndzbetween}(b) of the form $dN\!/dX = -0.28\,(\pm0.04) z + 0.92\,(\pm0.04)$, including the lowest redshift data point taken from~\cite{Narayanan2005}. This function shows $dN\!/dX$ going to zero at $z = 3.3$, implying that no ${\MgII}$ absorbers with equivalent widths between $0.02 < W_{r}^{\lambda2796} \le 0.3$~{\AA} should be detected above this redshift. In other words, weak ${\MgII}$ absorbers are predicted not to exist at redshifts above $z = 3.3$ by this trend. This evolution follows similar trends to the evolution in cosmic metallicity and ionizing background intensity, such that we detect more weak ${\MgII}$ absorbers at low redshift, where the metallicity of the CGM is higher and the ionizing background is less likely to destroy ${\MgII}$, as opposed to higher redshifts, where the opposite applies.

\cite{Steidel1992}, and later \cite{Nestor2005}, examined the redshift evolution of $dN\!/dz$ for strong ${\MgII}$ absorbers with $W_r^{\lambda2796} > 0.3$~{\AA}. They found that the number of strong ${\MgII}$ absorbers per redshift path length increases as a function of redshift from $z = 0$ to $z = 2.2$; however, they could not derive the slope of this trend to sufficient accuracy to distinguish between an evolving population or a non-evolving population. We perform a similar analysis on our sample, calculating instead $dN\!/dX$, where a flat distribution implies no evolution. When we take absorbers with $W_r^{\lambda2796} > 0.3$~{\AA}, we find that a fit to the function $dN\,/dX = \frac{c}{H_o}n_0\,\sigma_0(1+z)^{\epsilon}$ with a slope of $\epsilon = -0.20 \pm 0.22$ is appropriate, implying that the comoving number density and/or cross-section of strong ${\MgII}$ absorbers does not significantly evolve. However, when considering even stronger absorbers with $W_r^{\lambda2796} > 1$~{\AA}, we do observe evolution in $dN\!/dX$, with the evolution parameter, $\epsilon$, becomming positive and increasing, as shown in Figure~\ref{fig:nsigmaepsilon}. This evolution does not seem to be influenced by either an increase in the cosmic metallicity at low redshifts, or a decrease in the intensity of the cosmic ionizing background near the present epoch, which should produce more favorable conditions for ${\MgII}$ absorbing gas. In fact, in order to seemingly oppose these factors, we must infer an increase in the total quantity of ${\MgII}$ absorbing gas, observed as high equivalent width ${\MgII}$ systems, outside galaxies near $z = 2$. In Section~\ref{sec:trendcauses}, we will explore the physical processes responsible for evolution in the universal distribution of ${\MgII}$ absorbing gas.

\begin{figure}[bth]
\epsscale{1.2}
\plotone{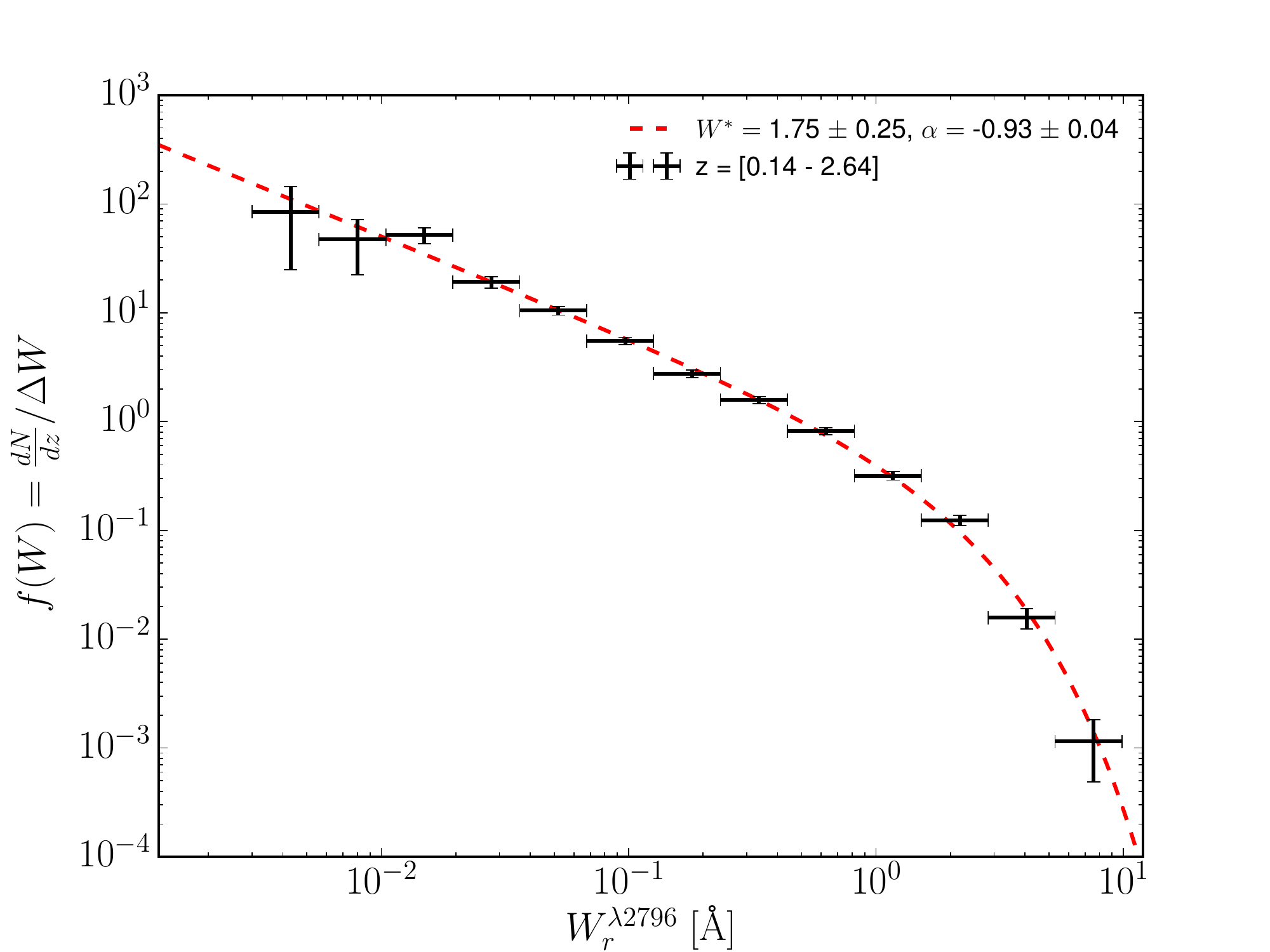}
\caption{The equivalent width distribution of all detected ${\MgII}$ absorbers, defined as the redshift path density ($dN\!/dz$) in each equivalent width bin divided by the bin width. The cosmic distribution of absorbing ${\MgII}$ is well fit by a Schechter function, with the parameters detailed in Equation~\ref{eqn:schechter}.}
\label{fig:totalewdistro}
\end{figure}

\cite{Kacprzak2011MgII} combined multiple previous studies to characterize the equivalent width distribution function, $f(W_r)$. It is important to note, for comparison, that their sample with $W_r < 0.3$~{\AA} spans redshifts from $0.4 \le z \le 1.4$, while their sample with $W_r \ge 0.3$~{\AA} spans $0.4 \le z \le 2.3$. They found a Schechter function with a low equivalent width slope of $\alpha = -0.642 \pm 0.062$ and a characteristic equivalent width for the exponential cutoff of $W^* = 0.97 \pm 0.06$~{\AA} best fit the data, within the reported $1\sigma$ uncertainties. In Figure~\ref{fig:totalewdistro}, we perform the same analysis with the total sample of our survey, finding $\alpha = -0.91 \pm 0.04$ and $W^* = 1.69 \pm 0.26$~{\AA}. The point of tension between these studies lies in the value of the low equivalent width slope. \cite{Kacprzak2011MgII} combine very different surveys in order to sample the full equivalent width distribution, leading to a non-uniform data set. Specifically at the low equivalent end of the distribution, we note that the error in their determination of $\alpha$ is smaller than the scatter in the data. In private communication with the authors, they speculate that differences in the analysis methods of low equivalent width ${\MgII}$ absorption line surveys lead to unaccounted systematics and an underestimation of the error in their functional fits.

At redshifts beyond $z \sim 2.5$, the works of~\cite{Matejek2012},~\cite{Matejek2013}, and~\cite{Chen2016} described the detection statistics of strong ${\MgII}$ absorbers up to $z = 7$ using the Magellan/FIRE spectrograph. They found that $dN\,/dX$ for all absorbers with $W_r^{\lambda2796} > 0.3$~{\AA} does not evolve from $z = 0.25$ to $z = 7$. However, restricting their sample to the absorbers with $W_r^{\lambda2796} > 1.0$~{\AA}, they found approximately a factor of 2 increase in number density at $z = 2-3$ compared to $z = 0$. At redshifts above $z = 3$, $dN\!/dX$ declines by an order of magnitude by $z \sim 6$. Converting ${\MgII}$ equivalent widths into an effective contribution to the global star formation rate using the methods of~\cite{Menard2011} and comparing them to the total galaxy star formation rate density derived from observations of deep fields, they found exceptional agreement in the evolutionary trends of both distributions at all redshifts. The incidence of strong ${\MgII}$ absorbers at all redshifts is tied to star formation.

\subsection{Potential Causes for Trends}
\label{sec:trendcauses}

The most obvious conclusion to be drawn from our analysis of ${\MgII}$ absorbers in The Vulture Survey is that the physical conditions affecting ${\MgII}$ absorbers change as a function of redshift. When we compare the sample of ${\MgII}$ absorbers at $z \sim 0.5$ to the sample of absorbers at $z \sim 2$, we find the following statements are true of the higher redshift sample:

\begin{enumerate}
\item There are more strong ${\MgII}$ absorbers per redshift path length and per absorption path.
\item The faint end slope of the equivalent width and column density distributions is flatter.
\item The ``knee'' of the Shechter fit to the equivalent width and column density distributions lies at higher values of $W_r^{\lambda2796}$ and $N(\MgII)$.
\item The cosmic mass density of ${\MgII}$ is larger.
\end{enumerate}

\noindent We can now state that the physical properties driving the global distribution of ${\MgII}$ absorbers at $z \sim 0.5$ are different than at $z \sim 2$. In addition, these physical properties do not affect all ${\MgII}$ systems equally. If we examine weak systems alone, we find that these low equivalent width absorbers follow a no-evolution expectation from present day up to $z \sim 1$, at which point they decrease in number per absorption path length rapidly until $z \sim 2.5$. If we examine the strongest systems, those with $W_r^{\lambda2796} > 1.0$~{\AA}, we find the opposite behavior, where these absorbers are more numerous at $z \sim 2.5$ than at $z \sim 0.5$. Possible explanations relate to the ionization conditions in the halos of galaxies at this time, the metallicity of gas around galaxies, and/or the quantity of metals in the circumgalactic medium, which we will now explore.

% H & M discussion
\cite{Haardt2012} represents the most recent estimate of the cosmic ionizing background as a function of redshift, which is the primary ionizing component responsible for the universal ionization state of gas in galactic halos. The authors found that the comoving 1 Ryd emissivity is an order of magnitude higher at $z = 1$ compared to the present epoch, and nearly two orders of magnitude larger at $z = 2$ than at present. As the number of ionizing photons in the IGM increases with increasing redshift, holding constant the density and quantity of metals in galactic halos, we would nominally expect for the ionization parameter of absorbers in the halos of galaxies to increase with redshift. Increasing the ionization parameter alone should decrease the observed quantity of ${\MgII}$, as it favors lower ionization parameter conditions. The comoving line density ($dN\!/dX$) of ${\MgII}$ absorbers with equivalent widths below $W_r^{\lambda2796} < 0.3$~{\AA} declines linearly with increasing redshift, following the expectation that a more intense ionizing background at higher redshift disfavors ${\MgII}$ absorption. However, when examining absorbers with $W_r^{\lambda2796} > 0.3$~{\AA}, we observe the opposite trend. These strong systems are not less numerous at higher redshift as a result of harsh ionizing conditions in the halos of galaxies. We therefore disfavor the hypothesis that changes in ionzation conditions in the halos of galaxies could drive the observed enhancement in the number of strong ${\MgII}$ absorbers at redshift $z \sim 2$ compared to $z \sim 0.5$.

% Metallicity discussion
The metallicity of the CGM has been best characterized by studies of DLAs and sub-DLAs located in the halos of galaxies, where many metal line transitions are observable, such as~\cite{Rafelski2012},~\cite{Quiret2016}, and~\cite{Jorgenson2013}. They found a rough trend of decreasing metallicity with increasing redshift dominated by scatter. These high column density systems certainly do not trace all ionized gas in the halos of galaxies as DLAs are predominantly neutral. Using a sample of gravitationally lensed galaxies, it has also been shown that the overall gas phase metallicity of star forming galaxies is 0.35 dex lower at $z \sim 2$ than at $z \sim 0$~\citep{Yuan2013}. In addition,~\cite{Wotta2016} and~\cite{Lehner2016} examined Lyman Limit Systems (LLS) and found that at $0.1 \le z \le 1.1$, the metallicity distribution is bimodal with peaks at $[X/H] \simeq -1.8$ and $-0.3$. At $z \ge 2$, they found a unimodal distribution with a peak at $[X/H] \simeq -2$. Assuming, then, that over time the overall metallicity of the circumgalactic medium increases, it would not be expected to observe larger quantities of ${\MgII}$ at $z \sim 2$ compared to $z \sim 0$. In fact, metals should build up over time in the halos of galaxies, producing more ${\MgII}$ absorption at lower redshift. We can better understand the role of metallicity by again comparing the sample of weak ${\MgII}$ absorbers to strong ${\MgII}$ absorbers. The weak systems follow the trend of rising CGM metallicity with time, by having a higher $dN\!/dX$ at $z \sim 0$ as compared to $z \sim 2$. However, the strongest ${\MgII}$ systems evolve in an opposite manner to trends in cosmic metallicity, implying that other physical properties drive their evolution. Furthermore, these very high equivalent width systems have the highest column densities, and with these strong systems being more numerous at higher redshifts, they drive the cosmic matter density of ${\MgII}$ absorbing gas to its highest value at $z \sim 2$. Cosmic metallicity evolution alone, then, cannot explain the slightly enhanced cosmic mass density of ${\MgII}$ absorbing gas at $z \sim 2$ as compared to $z \sim 0$.

% More stuff in halos
This leaves us, then, with the most likely conclusion being that galaxies eject more gas into their halos at $z \sim 2$ than at any other time in the form of strong, high equivalent width absorbers. \cite{Behroozi2013sfr} combined galaxy star formation rate measurements from 19 independent studies from $2006-2012$, spanning redshifts from $0 < z < 8$, to examine evolution in the cosmic star formation rate. They found that galaxies at $z \sim 2$ were forming stars at higher rates than any other time in cosmic history. It follows then that galaxies were ejecting more metal-enriched gas at $z \sim 2$ than at present through supernovae-driven outflows. Under the assumption that these outflows are observable as ${\MgII}$ absorption systems, they should manifenst themselves in the form of high equivalent width, high column density ${\MgII}$ absorbers, dense enough to be shielded from the more intense ionizing background at $z \sim 2$, enhanced in metallicity due to their supernovae origins. Therefore, despite competing factors such as a more intense ionizing background and lower overall cosmic metallicity at higher redshifts, we should measure a larger value of $\Omega_{\hbox{\scriptsize {\MgII}}}$ at $z \sim 2$ than at present as a result of galaxy feedback processes ejecting large quantities of ${\MgII}$ absorbing gas into their halos.

% What are weak MgII absorbers, then, if strongs are winds?
Strong ${\MgII}$ absorbers still exist at low redshift---their comoving line density does not go to zero at any point in time. However, at low redshifts near the present epoch, the majority of ${\MgII}$ absorbers are weak, low equivalent width, low column density systems whose properties are likely governed more by metallicity and incident radiation, due to their lower densities disfavoring any self-shielding. These weak absorbers may represent a population of either fragmented outflows, or newly condensed gas in the halos of galaxies~\citep{Maller2004}. Their eventual fate may be to accrete onto their host galaxy, but this rate of accretion must be small compared to the rate at which the cosmic ionizing background weakens and the cosmic metallicity of the halo increases towards lower redshifts. In other words, the rate of creation of weak ${\MgII}$ absorbers over time from $z \sim 2$ to the present epoch is greater than the rate of destruction, and this rate of creation may not be tied to an increased quantity of ${\MgII}$ absorbing gas in the halos of galaxies, but instead an evolution in cosmic properties governing the ionization state of all gas in galactic halos. We must now ask why the weakest systems build up slowly from $z > 2$ to the present epoch, directly opposing the trends observed for the strongest systems. To answer this question, we must understand the innate physical similarities and differences between high equivalent, high column density absorbers and low equivalent width, low column density absorbers.

% Stuff's within the Virial Radius
Studies associated with COS-Halos~\citep{Tumlinson2011} seeking to understand the distribution of metals around galaxies have found the majority of cool, metal absorbing gas lies within the virial radius of galaxies~\citep{Peeples2014}. \cite{Stern2016} also found that the mean cool gas density profile around galaxies scales as $R^{-1}$, with most strong, low ionization metal absorbers existing near the galaxy itself. Further refining these assertions specifically for ${\MgII}$,~\cite{Churchill2013letter} found that the covering fraction of ${\MgII}$ absorbers falls to $0$ at projected distances beyond the virial radius. This means that the ${\MgII}$ absorbers observed in The Vulture Survey likely lie within the virial radius of a galaxy. The ${\MgII}$ absorbing gas observed near $z \sim 2$ should eventually fall onto its host galaxy, or stall at large radii and remain a permanent part of the host galaxy halo~\citep{Oppenheimer2010,Ford2014}. What we observe in the form of low redshift ${\MgII}$ absorbers therefore may not only be the result of outflows near the present epoch, but also the remnants of the most energetic star-forming period in cosmic history.

% Strong absorbers are wind-related.
Perhaps the most clear-cut evidence that many very strong ($W_r^{\lambda2796} > 1.0$~{\AA}) absorbers have their origins in star formation driven winds comes from the works of~\cite{Matejek2012} and~\cite{Chen2016}. The authors examined their sample of high equivalent width ${\MgII}$ absorbers spanning redshifts $2 < z < 7$, combined with samples from~\cite{Nestor2005} and~\cite{Prochter2006} with data below $z < 2$, to relate the ${\MgII}$ population distribution with the cosmic star formation rate density. Following the methods of~\cite{Menard2011}, they convert the ${\MgII}$ rest frame equivalent widths to a comoving ${\OII}$ luminosity density, and from there derive a star formation rate density. When these ${\MgII}$-derived star formation rate densities were compared with the mean galaxy specific star formation rates from $0 < z < 7$, they found remarkable agreement. We therefore corroborate their results, finding good agreement between the evolutionary behavior of $dN\,/dX$ for strong ${\MgII}$ absorbers and the cosmic star formation rate density. These high equivalent width ${\MgII}$ systems are preferentially produced by feedback processes associated with star formation.

% Number of weak MgII absorbers at low z is increasing, not the cross-section
Referring back to Figure~\ref{fig:nsigmaepsilon}(a), we can examine how $\frac{c}{H_o}n_0\,\sigma_0$ changes for absorbers of different equivalent widths. Here, we can infer that the physical parameter causing $dN\!/dX$ to fall between $z \sim 0$ to $z \sim 2$ is the number density of weak ${\MgII}$ absorbers in galaxy halos ($n_0$), instead of the absorbing cross-section ($\sigma_0$). In~\cite{Evans2013}, the authors use Cloudy~\citep{Ferland2013} to estimate the size and number densities of clouds giving rise to ${\MgII}$ absorption. According to these simulations, absorber sizes are roughly an order of magnitude smaller at $z \sim 0$ than they were at $z \sim 2$ for clouds of the same hydrogen density across a wide range of densities. This implies the absorbing cross-section is lower at present than at $z \sim 2$, and, therefore, the number density of weak ${\MgII}$ absorbers should be the driving factor leading to the enhancement of $dN\!/dX$ for low equivalent width systems at the present epoch.

% Cooling radius discussion
Furthermore,~\cite{MAGIICAT3} discussed how the theoretical cooling radius, the radial distance from the center of a halo at which the initial gas density equals the characteristic density at which gas can cool, may serve as an important factor in determining the observational properties of ${\MgII}$ absorbers. They found that within the cooling radius, the covering fraction of ${\MgII}$ absorption increases, and does so most dramatically for absorbers with equivalent widths below $W_r^{\lambda2796} = 0.3$~{\AA}. At equivalent widths above this level, covering fractions inside and outside the cooling radius are statistically similar. This points toward a possible physical origin for a substantial portion of weak, low equivalent width ${\MgII}$ absorbers---these clouds cool and condense from collapsing gas clouds within the cooling radius. This gas could have originated from fragmented outflows or infalling material which, over time, met the conditions required to cool enough to be observed as ${\MgII}$, instead of a higher, more ionized species.

% The big picture
Given the above considerations, we favor a picture where galaxies, during the most active epoch of star formation in cosmic history, expell large quantities of metal enriched gas into their halos through star formation driven outflows at $z \sim 2$. These outflows manifest themselves in observations as strong ${\MgII}$ absorbers. Their destiny will be to eventually either fall back onto the galaxy and enrich the ISM or remain in the halo, subject to the cosmic ionizing background and possible fragmentation. Those absorbers which remain intact and accrete onto the galaxy are lost from observations at lower redshifts, while those which remain at larger radii will likely reach equilibrium with the ionization conditions in the halo. As the intensity of the cosmic ionizing background decreases with time, and more metals are ejected into the halo over time, more of this gas appears as weak ${\MgII}$ absorption near $z \sim 0$. Many of these weak ${\MgII}$ absorbers are likely more passive in nature, participating less in the overall baryon cycle in and out of galaxies, subject primarily to the environmental conditions of the CGM.

% ================ CONCLUSIONS ================
\section{Conclusions}
\label{sec:conclusions}

Using archival data from VLT/UVES and Keck/HIRES, we have undertaken the most complete survey of ${\MgII}$ absorbers in 602 quasar spectra with high resolution ($\sim 7~\mathrm{\kms}$), allowing for the detection of both strong and weak {\MgII} absorbsers. Our survey spans absorption redshifts from $0.18 < z < 2.57$, allowing for characterization of the evolution of the distribution of these absorbers across cosmic time. Using our own detection and analysis software, we are able to accurately characterize the equivalent width detection limit, absorption path length, and survey copmleteness to a level allowing for an accurate determination of $dN\!/dz$, the equivalent width distribution function, the column density distribution function, and the total cosmic mass density of {\MgII} absorbers. Our main findings are as follows:

\begin{enumerate}
\item We find 1180 intervening ${\MgII}$ absorption line systems with equivalent widths from $0.003$~{\AA} to $8.5$~{\AA}, and redshifts spanning $0.14 \le z \le 2.64$.
\item We present the distributions of the number of absorbers per unit redshift, $dN\!/dz$, and the comoving ${\MgII}$ line density, $dN\!/dX$, as a function of minimum equivalent width threshold and as a function of redshift. We parameterize the evolution in $dN\!/dX$ specifically with an emperical fit to the distribution in the form of $dN\,/dX = \frac{c}{H_o}n_0\,\sigma_0(1 + z)^{\epsilon}$, showing $\epsilon$ increases from $\epsilon=-1.11$ for all absorbers with $W_r^{\lambda2796} \ge 0.01$~{\AA} to $\epsilon=0.88$ for absorbers with $W_r^{\lambda2796} \ge 2$~{\AA}. High equivalent width ${\MgII}$ absorbers decrease in relative number per absorption path length from $z = 2$ to present, and low equivalent width ${\MgII}$ absorbers increase in relative number per absorption path length from $z = 2$ to present. We observe no evolution with redshift in absorbers with equivalent widths between $0.2 < W_r^{\lambda2796} < 1$~{\AA}.
\item We derive a closed form analytic parameterization of $dN\!/dX$ for all ${\MgII}$ absorbers. As shown in Equation~\ref{eqn:dndxanalytic}, $dN\!/dX$ can be expressed as a function of $W_{r,\mathrm{min}}^{\lambda2796}$ and $z$. In this parameterization, we also show how the comoving Hubble optical depth, $\frac{c}{H_o}n_0\,\sigma_0$, and evolution parameter, $\epsilon$ evolve with increasing minimum sample equivalent width (see Figure~\ref{fig:nsigmaepsilon}). These form a physical basis for understanding how the comoving line density of ${\MgII}$ absorbers evolves over cosmic time.
\item The equivalent width distribution function and the column density distribution function for ${\MgII}$ absorbers are both well fit by a Schecter Function, with a characteristic normalization, faint end slope, and exponential cutoff. Both functions show redshift evolution, specifically in the faint end slope, with this slope becoming shallower for redshifts near $z \sim 2$ as compared to the present epoch. There exist proportionately more high equivalent width, high column density ${\MgII}$ absorbers near $z \sim 2$ than at present.
\item The cosmic mass density of ${\MgII}$ absorbing gas, $\Omega_{\hbox{\scriptsize {\MgII}}}$, increases from $\Omega_{\hbox{\scriptsize {\MgII}}} \simeq 0.8\times10^{-8}$ at $z = 0.5$ to $\Omega_{\hbox{\scriptsize {\MgII}}} \simeq 1.3\times10^{-8}$ at $z \sim 2$.
\item The evolution in $dN\,/dX$ for the highest equivalent width ${\MgII}$ absorbers ($W_r^{\lambda2796} > 1.0$~{\AA}) follows evolutionary trends in the cosmic star formation rate density, which peaks near $z \sim 2$ and falls towards $z \sim 0$. This implies a connection between these very strong systems and star formation driven outflows. Examining other possible factors which could influence the properties of ${\MgII}$ absorbing gas, such as evolution in the intensity of the cosmic ionizing background and changes in cosmic metallicity, we find that the primary effect driving the evolution of the strongest ${\MgII}$ absorbers is the quantity of metal enriched gas expelled through star formation driven feedback.
\item We interpret the evolution of low equivalent width ${\MgII}$ absorbers ($W_r^{\lambda2796} < 0.3$~{\AA}) as a natural consequence of the evolution in the cosmic ionizing background and the metallicity of galaxy halos. These weaker systems potentially originate in the fragmented remains of star formation driven outflows and lower density gas clouds condensing within the host galaxy's cooling radius. They likely represent a lower density, passive component of the CGM, physically distinct from the strongest ${\MgII}$ systems.
\end{enumerate}

M.T.M. thanks the Australian Research Council for \textsl{Discovery Project} grant DP130100568 which supported this work. C.W.C. thanks the National Science Foundation for the grant AST-1517816, which partially supported this work.

\bibliographystyle{apj}
\bibliography{bibliography}

\begin{thebibliography}{77}
\expandafter\ifx\csname natexlab\endcsname\relax\def\natexlab#1{#1}\fi

\bibitem[{{Bagdonaite} {et~al.}(2014){Bagdonaite}, {Ubachs}, {Murphy}, \&
  {Whitmore}}]{Bagdonaite2014}
{Bagdonaite}, J., {Ubachs}, W., {Murphy}, M.~T., \& {Whitmore}, J.~B. 2014,
  \apj, 782, 10

\bibitem[{{Behroozi} {et~al.}(2013){Behroozi}, {Wechsler}, \&
  {Conroy}}]{Behroozi2013sfr}
{Behroozi}, P.~S., {Wechsler}, R.~H., \& {Conroy}, C. 2013, \apj, 770, 57

\bibitem[{{Bordoloi} {et~al.}(2014){Bordoloi}, {Lilly}, {Kacprzak}, \&
  {Churchill}}]{Bordoloi2014}
{Bordoloi}, R., {Lilly}, S.~J., {Kacprzak}, G.~G., \& {Churchill}, C.~W. 2014,
  \apj, 784, 108

\bibitem[{{Bordoloi} {et~al.}(2011){Bordoloi}, {Lilly}, {Knobel}, {Bolzonella},
  {Kampczyk}, {Carollo}, {Iovino}, {Zucca}, {Contini}, {Kneib}, {Le Fevre},
  {Mainieri}, {Renzini}, {Scodeggio}, {Zamorani}, {Balestra}, {Bardelli},
  {Bongiorno}, {Caputi}, {Cucciati}, {de la Torre}, {de Ravel}, {Garilli},
  {Kova{\v c}}, {Lamareille}, {Le Borgne}, {Le Brun}, {Maier}, {Mignoli},
  {Pello}, {Peng}, {Perez Montero}, {Presotto}, {Scarlata}, {Silverman},
  {Tanaka}, {Tasca}, {Tresse}, {Vergani}, {Barnes}, {Cappi}, {Cimatti},
  {Coppa}, {Diener}, {Franzetti}, {Koekemoer}, {L{\'o}pez-Sanjuan},
  {McCracken}, {Moresco}, {Nair}, {Oesch}, {Pozzetti}, \&
  {Welikala}}]{Bordoloi2011}
{Bordoloi}, R., {Lilly}, S.~J., {Knobel}, C., {et~al.} 2011, \apj, 743, 10

\bibitem[{{Bouch{\'e}} {et~al.}(2006){Bouch{\'e}}, {Murphy}, {P{\'e}roux},
  {Csabai}, \& {Wild}}]{Bouche2006}
{Bouch{\'e}}, N., {Murphy}, M.~T., {P{\'e}roux}, C., {Csabai}, I., \& {Wild},
  V. 2006, \mnras, 371, 495

\bibitem[{{Charlton} {et~al.}(2000){Charlton}, {Mellon}, {Rigby}, \&
  {Churchill}}]{Charlton2000}
{Charlton}, J.~C., {Mellon}, R.~R., {Rigby}, J.~R., \& {Churchill}, C.~W. 2000,
  \apj, 545, 635

\bibitem[{{Chen} {et~al.}(2010){Chen}, {Wild}, {Tinker}, {Gauthier}, {Helsby},
  {Shectman}, \& {Thompson}}]{Chen2010b}
{Chen}, H.-W., {Wild}, V., {Tinker}, J.~L., {et~al.} 2010, \apjl, 724, L176

\bibitem[{{Chen} {et~al.}(2016){Chen}, {Simcoe}, {Torrey}, {Ba{\~n}ados},
  {Cooksey}, {Cooper}, {Furesz}, {Matejek}, {Miller}, {Turner}, {Venemans},
  {Decarli}, {Farina}, {Mazzucchelli}, \& {Walter}}]{Chen2016}
{Chen}, S.-F.~S., {Simcoe}, R.~A., {Torrey}, P., {et~al.} 2016, ArXiv e-prints

\bibitem[{{Churchill} {et~al.}(2013{\natexlab{a}}){Churchill}, {Nielsen},
  {Kacprzak}, \& {Trujillo-Gomez}}]{Churchill2013letter}
{Churchill}, C.~W., {Nielsen}, N.~M., {Kacprzak}, G.~G., \& {Trujillo-Gomez},
  S. 2013{\natexlab{a}}, \apjl, 763, L42

\bibitem[{{Churchill} {et~al.}(1999){Churchill}, {Rigby}, {Charlton}, \&
  {Vogt}}]{Churchill1999}
{Churchill}, C.~W., {Rigby}, J.~R., {Charlton}, J.~C., \& {Vogt}, S.~S. 1999,
  \apjs, 120, 51

\bibitem[{{Churchill} {et~al.}(2013{\natexlab{b}}){Churchill},
  {Trujillo-Gomez}, {Nielsen}, \& {Kacprzak}}]{MAGIICAT3}
{Churchill}, C.~W., {Trujillo-Gomez}, S., {Nielsen}, N.~M., \& {Kacprzak},
  G.~G. 2013{\natexlab{b}}, \apj, 779, 87

\bibitem[{{Churchill} \& {Vogt}(2001)}]{Churchill2001}
{Churchill}, C.~W., \& {Vogt}, S.~S. 2001, \aj, 122, 679

\bibitem[{{Churchill} {et~al.}(2003){Churchill}, {Vogt}, \&
  {Charlton}}]{Churchill2003}
{Churchill}, C.~W., {Vogt}, S.~S., \& {Charlton}, J.~C. 2003, \aj, 125, 98

\bibitem[{Davidson \& MacKinnon(2000)}]{Davidson2000bootstrap}
Davidson, R., \& MacKinnon, J.~G. 2000, Econometric Reviews, 19, 55

\bibitem[{{Dekker} {et~al.}(2000){Dekker}, {D'Odorico}, {Kaufer}, {Delabre}, \&
  {Kotzlowski}}]{Dekker2000}
{Dekker}, H., {D'Odorico}, S., {Kaufer}, A., {Delabre}, B., \& {Kotzlowski}, H.
  2000, in \procspie, Vol. 4008, Optical and IR Telescope Instrumentation and
  Detectors, ed. M.~{Iye} \& A.~F. {Moorwood}, 534--545

\bibitem[{{Evans} {et~al.}(2013){Evans}, {Churchill}, {Murphy}, {Nielsen}, \&
  {Klimek}}]{Evans2013}
{Evans}, J.~L., {Churchill}, C.~W., {Murphy}, M.~T., {Nielsen}, N.~M., \&
  {Klimek}, E.~S. 2013, \apj, 768, 3

\bibitem[{{Ferland} {et~al.}(2013){Ferland}, {Porter}, {van Hoof}, {Williams},
  {Abel}, {Lykins}, {Shaw}, {Henney}, \& {Stancil}}]{Ferland2013}
{Ferland}, G.~J., {Porter}, R.~L., {van Hoof}, P.~A.~M., {et~al.} 2013, Revista
  Mexicana de Astronomia y Astrofisica, 49, 137

\bibitem[{{Ford} {et~al.}(2014){Ford}, {Dav{\'e}}, {Oppenheimer}, {Katz},
  {Kollmeier}, {Thompson}, \& {Weinberg}}]{Ford2014}
{Ford}, A.~B., {Dav{\'e}}, R., {Oppenheimer}, B.~D., {et~al.} 2014, \mnras,
  444, 1260

\bibitem[{{Ford} {et~al.}(2013){Ford}, {Oppenheimer}, {Dav{\'e}}, {Katz},
  {Kollmeier}, \& {Weinberg}}]{Ford2013mass}
{Ford}, A.~B., {Oppenheimer}, B.~D., {Dav{\'e}}, R., {et~al.} 2013, \mnras,
  432, 89

\bibitem[{{Gauthier} {et~al.}(2009){Gauthier}, {Chen}, \&
  {Tinker}}]{Gauthier2009}
{Gauthier}, J.-R., {Chen}, H.-W., \& {Tinker}, J.~L. 2009, \apj, 702, 50

\bibitem[{{Haardt} \& {Madau}(2012)}]{Haardt2012}
{Haardt}, F., \& {Madau}, P. 2012, \apj, 746, 125

\bibitem[{{Jorgenson} {et~al.}(2013){Jorgenson}, {Murphy}, \&
  {Thompson}}]{Jorgenson2013}
{Jorgenson}, R.~A., {Murphy}, M.~T., \& {Thompson}, R. 2013, \mnras, 435, 482

\bibitem[{{Kacprzak} \& {Churchill}(2011)}]{Kacprzak2011MgII}
{Kacprzak}, G.~G., \& {Churchill}, C.~W. 2011, \apjl, 743, L34

\bibitem[{{Kacprzak} {et~al.}(2011){Kacprzak}, {Churchill}, {Evans}, {Murphy},
  \& {Steidel}}]{Kacprzak2011}
{Kacprzak}, G.~G., {Churchill}, C.~W., {Evans}, J.~L., {Murphy}, M.~T., \&
  {Steidel}, C.~C. 2011, \mnras, 416, 3118

\bibitem[{{Kacprzak} {et~al.}(2012){Kacprzak}, {Churchill}, \&
  {Nielsen}}]{Kacprzak2012-PA}
{Kacprzak}, G.~G., {Churchill}, C.~W., \& {Nielsen}, N.~M. 2012, \apjl, 760, L7

\bibitem[{{King} {et~al.}(2012){King}, {Webb}, {Murphy}, {Flambaum},
  {Carswell}, {Bainbridge}, {Wilczynska}, \& {Koch}}]{King2012}
{King}, J.~A., {Webb}, J.~K., {Murphy}, M.~T., {et~al.} 2012, \mnras, 422, 3370

\bibitem[{{Kulkarni} \& {Fall}(2002)}]{Kulkarni2002}
{Kulkarni}, V.~P., \& {Fall}, S.~M. 2002, \apj, 580, 732

\bibitem[{{Kulkarni} {et~al.}(2005){Kulkarni}, {Fall}, {Lauroesch}, {York},
  {Welty}, {Khare}, \& {Truran}}]{Kulkarni2005}
{Kulkarni}, V.~P., {Fall}, S.~M., {Lauroesch}, J.~T., {et~al.} 2005, \apj, 618,
  68

\bibitem[{{Kulkarni} {et~al.}(2007){Kulkarni}, {Khare}, {P{\'e}roux}, {York},
  {Lauroesch}, \& {Meiring}}]{Kulkarni2007}
{Kulkarni}, V.~P., {Khare}, P., {P{\'e}roux}, C., {et~al.} 2007, \apj, 661, 88

\bibitem[{{Lanzetta} {et~al.}(1987){Lanzetta}, {Turnshek}, \&
  {Wolfe}}]{Lanzetta1987}
{Lanzetta}, K.~M., {Turnshek}, D.~A., \& {Wolfe}, A.~M. 1987, \apj, 322, 739

\bibitem[{{Lehner} {et~al.}(2016){Lehner}, {O'Meara}, {Howk}, {Prochaska}, \&
  {Fumagalli}}]{Lehner2016}
{Lehner}, N., {O'Meara}, J.~M., {Howk}, J.~C., {Prochaska}, J.~X., \&
  {Fumagalli}, M. 2016, ArXiv e-prints

\bibitem[{{Lovegrove} \& {Simcoe}(2011)}]{Lovegrove2011}
{Lovegrove}, E., \& {Simcoe}, R.~A. 2011, \apj, 740, 30

\bibitem[{{Lundgren} {et~al.}(2009){Lundgren}, {Brunner}, {York}, {Ross},
  {Quashnock}, {Myers}, {Schneider}, {Al Sayyad}, \& {Bahcall}}]{Lundgren2009}
{Lundgren}, B.~F., {Brunner}, R.~J., {York}, D.~G., {et~al.} 2009, \apj, 698,
  819

\bibitem[{{Maller} \& {Bullock}(2004)}]{Maller2004}
{Maller}, A.~H., \& {Bullock}, J.~S. 2004, \mnras, 355, 694

\bibitem[{{Martin} \& {Bouch{\'e}}(2009)}]{Martin2009}
{Martin}, C.~L., \& {Bouch{\'e}}, N. 2009, \apj, 703, 1394

\bibitem[{{Matejek} \& {Simcoe}(2012)}]{Matejek2012}
{Matejek}, M.~S., \& {Simcoe}, R.~A. 2012, \apj, 761, 112

\bibitem[{{Matejek} {et~al.}(2013){Matejek}, {Simcoe}, {Cooksey}, \&
  {Seyffert}}]{Matejek2013}
{Matejek}, M.~S., {Simcoe}, R.~A., {Cooksey}, K.~L., \& {Seyffert}, E.~N. 2013,
  \apj, 764, 9

\bibitem[{{M{\'e}nard} \& {Chelouche}(2009)}]{Menard2009}
{M{\'e}nard}, B., \& {Chelouche}, D. 2009, \mnras, 393, 808

\bibitem[{{M{\'e}nard} {et~al.}(2011){M{\'e}nard}, {Wild}, {Nestor}, {Quider},
  {Zibetti}, {Rao}, \& {Turnshek}}]{Menard2011}
{M{\'e}nard}, B., {Wild}, V., {Nestor}, D., {et~al.} 2011, \mnras, 417, 801

\bibitem[{{Murphy}(2016)}]{MurphyPOPLER}
{Murphy}, M.~T. 2016, {{UVES\_popler}: {POst} {PipeLine} {Echelle} {Reduction}
  software}, \mbox{doi}:\url{10.5281/zenodo.44765}

\bibitem[{{Murphy}(in prep)}]{Murphyprep}
---. in prep, \mnras

\bibitem[{{Murphy} {et~al.}(2016){Murphy}, {Malec}, \&
  {Prochaska}}]{Murphy2016}
{Murphy}, M.~T., {Malec}, A.~L., \& {Prochaska}, J.~X. 2016, \mnras, 461, 2461

\bibitem[{{Narayanan} {et~al.}(2005){Narayanan}, {Charlton}, {Masiero}, \&
  {Lynch}}]{Narayanan2005}
{Narayanan}, A., {Charlton}, J.~C., {Masiero}, J.~R., \& {Lynch}, R. 2005,
  \apj, 632, 92

\bibitem[{{Narayanan} {et~al.}(2007){Narayanan}, {Misawa}, {Charlton}, \&
  {Kim}}]{Narayanan2007}
{Narayanan}, A., {Misawa}, T., {Charlton}, J.~C., \& {Kim}, T.-S. 2007, \apj,
  660, 1093

\bibitem[{{Nestor} {et~al.}(2011){Nestor}, {Johnson}, {Wild}, {M{\'e}nard},
  {Turnshek}, {Rao}, \& {Pettini}}]{Nestor2011}
{Nestor}, D.~B., {Johnson}, B.~D., {Wild}, V., {et~al.} 2011, \mnras, 412, 1559

\bibitem[{{Nestor} {et~al.}(2005){Nestor}, {Turnshek}, \& {Rao}}]{Nestor2005}
{Nestor}, D.~B., {Turnshek}, D.~A., \& {Rao}, S.~M. 2005, \apj, 628, 637

\bibitem[{{Nielsen} {et~al.}(2013{\natexlab{a}}){Nielsen}, {Churchill}, \&
  {Kacprzak}}]{MAGIICAT2}
{Nielsen}, N.~M., {Churchill}, C.~W., \& {Kacprzak}, G.~G. 2013{\natexlab{a}},
  \apj, 776, 115

\bibitem[{{Nielsen} {et~al.}(2013{\natexlab{b}}){Nielsen}, {Churchill},
  {Kacprzak}, \& {Murphy}}]{MAGIICAT1}
{Nielsen}, N.~M., {Churchill}, C.~W., {Kacprzak}, G.~G., \& {Murphy}, M.~T.
  2013{\natexlab{b}}, \apj, 776, 114

\bibitem[{{Nielsen} {et~al.}(2015){Nielsen}, {Churchill}, {Kacprzak}, {Murphy},
  \& {Evans}}]{MAGIICAT5}
{Nielsen}, N.~M., {Churchill}, C.~W., {Kacprzak}, G.~G., {Murphy}, M.~T., \&
  {Evans}, J.~L. 2015, \apj, 812, 83

\bibitem[{{Nielsen} {et~al.}(2016){Nielsen}, {Churchill}, {Kacprzak}, {Murphy},
  \& {Evans}}]{MAGIICAT4}
---. 2016, \apj, 818, 171

\bibitem[{{Noterdaeme} {et~al.}(2010){Noterdaeme}, {Srianand}, \&
  {Mohan}}]{Noterdaeme2010}
{Noterdaeme}, P., {Srianand}, R., \& {Mohan}, V. 2010, \mnras, 403, 906

\bibitem[{{O'Meara} {et~al.}(2015){O'Meara}, {Lehner}, {Howk}, {Prochaska},
  {Fox}, {Swain}, {Gelino}, {Berriman}, \& {Tran}}]{OMeara2015}
{O'Meara}, J.~M., {Lehner}, N., {Howk}, J.~C., {et~al.} 2015, \aj, 150, 111

\bibitem[{{Oppenheimer} {et~al.}(2010){Oppenheimer}, {Dav{\'e}}, {Kere{\v s}},
  {Fardal}, {Katz}, {Kollmeier}, \& {Weinberg}}]{Oppenheimer2010}
{Oppenheimer}, B.~D., {Dav{\'e}}, R., {Kere{\v s}}, D., {et~al.} 2010, \mnras,
  406, 2325

\bibitem[{{Peeples} {et~al.}(2014){Peeples}, {Werk}, {Tumlinson},
  {Oppenheimer}, {Prochaska}, {Katz}, \& {Weinberg}}]{Peeples2014}
{Peeples}, M.~S., {Werk}, J.~K., {Tumlinson}, J., {et~al.} 2014, \apj, 786, 54

\bibitem[{{Planck Collaboration} {et~al.}(2016){Planck Collaboration}, {Ade},
  {Aghanim}, {Arnaud}, {Ashdown}, {Aumont}, {Baccigalupi}, {Banday},
  {Barreiro}, {Bartlett}, \& et~al.}]{Planck2016}
{Planck Collaboration}, {Ade}, P.~A.~R., {Aghanim}, N., {et~al.} 2016, \aap,
  594, A13

\bibitem[{{Prochaska} {et~al.}(2003){Prochaska}, {Gawiser}, {Wolfe}, {Castro},
  \& {Djorgovski}}]{Prochaska2003}
{Prochaska}, J.~X., {Gawiser}, E., {Wolfe}, A.~M., {Castro}, S., \&
  {Djorgovski}, S.~G. 2003, \apjl, 595, L9

\bibitem[{{Prochter} {et~al.}(2006){Prochter}, {Prochaska}, \&
  {Burles}}]{Prochter2006}
{Prochter}, G.~E., {Prochaska}, J.~X., \& {Burles}, S.~M. 2006, \apj, 639, 766

\bibitem[{{Quiret} {et~al.}(2016){Quiret}, {P{\'e}roux}, {Zafar}, {Kulkarni},
  {Jenkins}, {Milliard}, {Rahmani}, {Popping}, {Rao}, {Turnshek}, \&
  {Monier}}]{Quiret2016}
{Quiret}, S., {P{\'e}roux}, C., {Zafar}, T., {et~al.} 2016, \mnras

\bibitem[{{Rafelski} {et~al.}(2012){Rafelski}, {Wolfe}, {Prochaska},
  {Neeleman}, \& {Mendez}}]{Rafelski2012}
{Rafelski}, M., {Wolfe}, A.~M., {Prochaska}, J.~X., {Neeleman}, M., \&
  {Mendez}, A.~J. 2012, \apj, 755, 89

\bibitem[{{Rigby} {et~al.}(2002){Rigby}, {Charlton}, \&
  {Churchill}}]{Rigby2002}
{Rigby}, J.~R., {Charlton}, J.~C., \& {Churchill}, C.~W. 2002, \apj, 565, 743

\bibitem[{{Rubin} {et~al.}(2010){Rubin}, {Weiner}, {Koo}, {Martin},
  {Prochaska}, {Coil}, \& {Newman}}]{Rubin2010}
{Rubin}, K.~H.~R., {Weiner}, B.~J., {Koo}, D.~C., {et~al.} 2010, \apj, 719,
  1503

\bibitem[{{Sargent} {et~al.}(1988){Sargent}, {Steidel}, \&
  {Boksenberg}}]{Sargent1988}
{Sargent}, W.~L.~W., {Steidel}, C.~C., \& {Boksenberg}, A. 1988, \apj, 334, 22

\bibitem[{{Sharma} \& {Nath}(2013)}]{Sharma2013}
{Sharma}, M., \& {Nath}, B.~B. 2013, \apj, 763, 17

\bibitem[{{Shattow} {et~al.}(2015){Shattow}, {Croton}, \&
  {Bibiano}}]{Shattow2015}
{Shattow}, G.~M., {Croton}, D.~J., \& {Bibiano}, A. 2015, \mnras, 450, 2306

\bibitem[{{Steidel} \& {Sargent}(1992)}]{Steidel1992}
{Steidel}, C.~C., \& {Sargent}, W.~L.~W. 1992, \apjs, 80, 1

\bibitem[{{Stern} {et~al.}(2016){Stern}, {Hennawi}, {Prochaska}, \&
  {Werk}}]{Stern2016}
{Stern}, J., {Hennawi}, J.~F., {Prochaska}, J.~X., \& {Werk}, J.~K. 2016, ArXiv
  e-prints

\bibitem[{{Stewart} {et~al.}(2011){Stewart}, {Kaufmann}, {Bullock}, {Barton},
  {Maller}, {Diemand}, \& {Wadsley}}]{Stewart2011}
{Stewart}, K.~R., {Kaufmann}, T., {Bullock}, J.~S., {et~al.} 2011, \apj, 738,
  39

\bibitem[{{Tremonti} {et~al.}(2007){Tremonti}, {Moustakas}, \&
  {Diamond-Stanic}}]{Tremonti2007}
{Tremonti}, C.~A., {Moustakas}, J., \& {Diamond-Stanic}, A.~M. 2007, \apjl,
  663, L77

\bibitem[{{Tumlinson} {et~al.}(2011){Tumlinson}, {Thom}, {Werk}, {Prochaska},
  {Tripp}, {Weinberg}, {Peeples}, {O'Meara}, {Oppenheimer}, {Meiring}, {Katz},
  {Dav{\'e}}, {Ford}, \& {Sembach}}]{Tumlinson2011}
{Tumlinson}, J., {Thom}, C., {Werk}, J.~K., {et~al.} 2011, Science, 334, 948

\bibitem[{{Tytler} {et~al.}(1987){Tytler}, {Boksenberg}, {Sargent}, {Young}, \&
  {Kunth}}]{Tytler1987}
{Tytler}, D., {Boksenberg}, A., {Sargent}, W.~L.~W., {Young}, P., \& {Kunth},
  D. 1987, \apjs, 64, 667

\bibitem[{{Vogt} {et~al.}(1994){Vogt}, {Allen}, {Bigelow}, {Bresee}, {Brown},
  {Cantrall}, {Conrad}, {Couture}, {Delaney}, {Epps}, {Hilyard}, {Hilyard},
  {Horn}, {Jern}, {Kanto}, {Keane}, {Kibrick}, {Lewis}, {Osborne},
  {Pardeilhan}, {Pfister}, {Ricketts}, {Robinson}, {Stover}, {Tucker}, {Ward},
  \& {Wei}}]{Vogt1994}
{Vogt}, S.~S., {Allen}, S.~L., {Bigelow}, B.~C., {et~al.} 1994, in \procspie,
  Vol. 2198, Instrumentation in Astronomy VIII, ed. D.~L. {Crawford} \& E.~R.
  {Craine}, 362

\bibitem[{{Weiner} {et~al.}(2009){Weiner}, {Coil}, {Prochaska}, {Newman},
  {Cooper}, {Bundy}, {Conselice}, {Dutton}, {Faber}, {Koo}, {Lotz}, {Rieke}, \&
  {Rubin}}]{Weiner2009}
{Weiner}, B.~J., {Coil}, A.~L., {Prochaska}, J.~X., {et~al.} 2009, \apj, 692,
  187

\bibitem[{{Wotta} {et~al.}(2016){Wotta}, {Lehner}, {Howk}, {O'Meara}, \&
  {Prochaska}}]{Wotta2016}
{Wotta}, C.~B., {Lehner}, N., {Howk}, J.~C., {O'Meara}, J.~M., \& {Prochaska},
  J.~X. 2016, \apj, 831, 95

\bibitem[{{Yuan} {et~al.}(2013){Yuan}, {Kewley}, \& {Richard}}]{Yuan2013}
{Yuan}, T.-T., {Kewley}, L.~J., \& {Richard}, J. 2013, \apj, 763, 9

\bibitem[{{Zhu} {et~al.}(2015){Zhu}, {Comparat}, {Kneib}, {Delubac},
  {Raichoor}, {Dawson}, {Newman}, {Y{\`e}che}, {Zhou}, \&
  {Schneider}}]{Zhu2015}
{Zhu}, B.~G., {Comparat}, J., {Kneib}, J.-P., {et~al.} 2015, \apj, 815, 48

\bibitem[{{Zhu} \& {M{\'e}nard}(2013)}]{Zhu2013}
{Zhu}, G., \& {M{\'e}nard}, B. 2013, \apj, 770, 130

\bibitem[{{Zibetti} {et~al.}(2007){Zibetti}, {M{\'e}nard}, {Nestor}, {Quider},
  {Rao}, \& {Turnshek}}]{Zibetti2007}
{Zibetti}, S., {M{\'e}nard}, B., {Nestor}, D.~B., {et~al.} 2007, \apj, 658, 161

\end{thebibliography}

\end{document}